\documentclass{aa}  

\usepackage{revsymb}        
\usepackage{amsmath}        
\usepackage[usenames]{color}    
\usepackage{epsfig}
\usepackage{ulem}
\usepackage[latin1]{inputenc}
\usepackage{txfonts}
\usepackage{color}
\usepackage{lscape}
\usepackage{natbib}
\def\ni{\noindent }

\begin{document}

   \title{Rotational splittings for slow to moderate rotators:}

   \subtitle{Latitudinal dependency or higher order effects in $\Omega$?}

   \author{R-M. Ouazzani
          \inst{1},
          \and
          M-J. Goupil
          \inst{1} }

   \institute{LESIA, UMR8109, Universit\'e Pierre et Marie Curie, Universit\'e Denis Diderot, Observatoire de Paris, 92195 Meudon Cedex, France\\
              \email{rhita-maria.ouazzani@obspm.fr}
                     }

   \date{Received April 1, 2011;  accepted May 16, 2011}

% \abstract{}{}{}{}{} 
% 5 {} token are mandatory
 
  \abstract
   {The unprecedented photometric quality reached by the CoRoT and Kepler space missions opens new prospects for studying stellar rotation. Information about the rotation rate is contained on the one hand in the low frequency part of power spectra, where signatures of nonuniform surface rotation are expected, and on the second hand in the frequency splittings  induced by the internal rotation rate.}
   {We wish to figure out whether the differences between the seismic rotation period as determined by a mean rotational splitting, and the  rotation period measured from the low frequency peak in the Fourier spectrum -- observed for some of  CoRoT's targets -- can provide constraints on the rotation profile.}
   {For uniform moderate rotators,perturbative corrections to second order and third order in terms of the rotation angular velocity $\Omega$, must not be neglected.
 These effects, in particular, may mimic differential rotation. We apply our perturbation method to evaluate mode frequencies accurate up to $\Omega^3$ for uniform rotation. Effects of latitudinal dependence are calculated in the linear approximation. Numerical results were obtained for selected models of the upper and lower main sequence. For the latitudinal dependence, we adopted two types of rotation profile: one with rotation uniform in depth, and one with a solar-like tachocline.}
   {
%Deviations from the first order splitting for a uniformly rotating star can be due to both cubic order effects of rotation and to latitudinal differential rotation. 
In models of $\beta$ Cephei pulsators, upper main sequence stars, third order effects become comparable to that of a horizontal shear similar to the solar one at rotation rates well below the breakup values. 
%These nonlinear effects are strongly mode-dependent. 
We show how a clean signature of the latitudinal shear may be extracted. Our models of two CoRoT target HD 181906 and HD 181420, which are solar-like pulsators, represent lower main sequence objects. These are slow rotators and nonlinear effects in splittings are accordingly small. We use data on one low frequency peak and one splitting of a dipolar mode to constrain the rotation profile in HD 181420 and HD 181906.}
   {The relative influence of the two effects strongly depends on the type of the oscillation modes at stake in the star and on the magnitude of the rotation rate. Given mean rotational splitting and the frequency of a spot signature, it is possible to distinguish between the two hypothesis, and in the case of differential rotation in latitude, we propose a method to determine the type of rotation profile and a range of values for the shear.}


   \maketitle
%
%________________________________________________________________

\section{Introduction}

 The CoRoT and Kepler space-borne missions with their uninterrupted
observations spanning a long-time interval promise a wealth of data
suitable for studying stellar rotation. Information about internal
rotation is contained  in characteristic spacings (known as the
rotational splittings) which  appear in the power spectra of light
curves.  Mean values of the rotational splitting have already been
determined for a number of stars -HD 181420 \citep{Barban2009}, HD
49933 \citep{Benomar2009}, V1449 Aql \citep{Belkacem2009} and HD 181906 \citep{Garcia2009}-
observed by CoRoT.  These values yield some information about average
rotation rates sampled by modes detected in these objects. Since all
the data concern p-modes, the mean values mostly reflect
the rotation rate in the outer layers. To probe deeper layers, we need
splittings for gravity modes. In three Fourier spectra analyzed so far
(HD 181420, HD 181906 and V1449 Aql), low frequency peaks were found but were
 attributed to the effects of spots on the rotating stellar surfaces
\citep[for a complete review on spot modeling, see]
[]{Collier2002,Mosser2009}. Rotation periods deduced this
way were found to be different from those determined from the splittings.
This is not surprising.  The spots only give access to the surface
rotation rate at the latitude of their location, whereas the splittings yield a
mode-dependent mean value of the interior profile. 

The linear relation between the splittings and the rotation
rate, $\Omega$, follows from the first order perturbative
treatment of the Coriolis acceleration \citep{Ledoux1945}. At
moderate rotation rates, the perturbative formalism may still be applicable but we have to go beyond the first
order \citep[][]{Reese2006,Ouazzani2009,Suarez2010,Burke2011}.
\cite{Saio1981}, \cite{GT90} and \cite{DG92} (hereafter DG92) derived  oscillation frequencies including second order corrections in  $\Omega$. Such corrections arise from
the higher order effects of the Coriolis acceleration and the lowest
order effects of the centrifugal acceleration, which causes
distortion of the stellar structure.  The oscillation frequencies
 no longer  depend on $\Omega$ in a linear way but  linear
rotational splittings may still be recovered from frequency differences
between  prograde and retrograde modes of the same degree and
order. This simple property is lost when rotation couples modes
with close frequencies. The formalism  must then be modified (DG92,
Soufi  et al. 1998) and the recovery is more difficult. The cubic
effects in $\Omega$  make it still more complicated. The resulting
difference between  prograde and retrograde modes becomes
dependent on the mode's azimuthal order, $m$, in such a way that it may be
misinterpreted as the effects of a latitudinal dependence of the
rotation rate.

The question  arises whether in the presence of significant
nonlinear effects it is still possible to extract the values of
linear splittings, which provide integral constraints on
differential rotation in the interior. The next question which we ask in
this paper is what may be learnt by combining such constraints with
data on low frequency peaks that are attributed to spots and yield
information on the surface rotation rate. We expect that the answers
depend on the type of pulsator, and the characteristics of its
observed modes. Here, we specifically consider two very different
types of main sequence objects, $\beta$ Cephei and a solar-like
pulsators. In the first case, we are dealing with a massive star
where low order p- and g-modes are unstable, and in the second case
 with low-mass stars where high order p-modes are damped but stochastically driven.

The paper is organized as follows. Sect. 2 gives the basic
theoretical framework for this study. In Sect. 3, we focus on the
effects of cubic order and near degeneracy contributions to
pulsation frequencies and present numerical results for a selected
model of a $\beta$ Cephei star. Explicit expressions for the
rotational splitting in the case of latitude dependent rotation profile are
given in Sect. 4. Departures from a linear dependence on $m$
are then compared with those connected to the nonlinear effects in
$\Omega$. Prospects for disentangling these two effects
are discussed in Sect.5, using the case of a $\beta$ Cephei
star as an example. Sect. 6 is devoted to two CoRoT targets, the solar type stars HD 181906 and HD181420,
for which we combine the splitting with the low frequency peak and make
some inference on the rotation profiles.
 Sect. 7 is dedicated to conclusions.

\section{Perturbational treatment of uniform rotation: effects on pulsation frequencies}
In the presence of rotation, the centrifugal and Coriolis
accelerations come into play. The centrifugal force affects the
structure of the star and distorts its shape. The  resonant cavity
is changed and with that the oscillation frequencies. The Coriolis force
enters the equation of motion and affects the motion of the waves,
hence the frequencies of normal modes. As rotation breaks the spherical
symmetry, it lifts the frequency degeneracy, introducing a
dependence on the azimuthal order, $m$. Without rotation, mode
frequencies only depend on the radial order, $n$, and the angular
degree, $\ell$, and are $2\ell+1$-fold degenerated.

To first order in the rotation rate, $\Omega$, the normal mode
frequencies in the inertial frame are given by $\omega_{n,\ell,m} =
\omega_{n,\ell,0} + m \, \Omega \, \beta_{n,\ell}$. The explicit
form of the last coefficient is provided in Eq.(\ref{def_beta}). Here, we only note that
$C_{n,\ell} = 1-\beta_{n,\ell}$ is known as the Ledoux
constant \citep{Ledoux1945}.

 Characteristic spacings appear in the spectrum, such as the rotational splitting that we define here
 as:
\begin{equation}
\mathfrak{s}_{n,\ell,m} = {\omega_{n,\ell,m}-\omega_{n,\ell,-m}\, / \,2 m}
\label{Sm}
\end{equation}
This gives us a basis for determining the rotation rate.
From now on, we drop the subscripts $(n,\ell,m)$  for $\omega$ and the splitting, unless
there is an ambiguity. We also use $\sigma=\omega/\Omega_k$
(where $\Omega_k \, = \, \sqrt{\rm G \rm M / \rm R^3}$ is
the break-up frequency) instead of $\omega$.

\subsection{Equilibrium configuration}
We here consider the case of uniform rotation. 
The stationary equation of motion in an inertial frame of reference is:
\begin{equation}
\label{motion}
(\mathbf{\rm v_0 \cdot \nabla}) \mathbf{\rm v_0} = -\frac{\mathbf{\nabla} \rm P}{\rho} -\mathbf{\nabla} \phi = \mathbf{\rm F}
\end{equation}
Where in the left hand side, $\vec{\rm v_0} = \vec{\Omega} \wedge \vec{\rm r} = \Omega_0 \, \rm r \, \sin \theta \, \vec{\rm e_{\phi}}$ in the spherical basis is the velocity field due to rotation at the angular velocity $\Omega_0$, $\theta$ being the colatitude \citep[see e.g.][]{UNNO1989}.
$P$, $\rho$ and $\phi$ are the pressure, density and
gravitational potential, respectively. For a rotating star, the left
hand side corresponds to the centrifugal acceleration $\mathbf{\rm
F}=-\mathbf{\Omega} \times (\mathbf{\Omega} \times \mathbf{\rm r})$,
whose effect on the equilibrium structure is twofold : on one hand, a spherically symmetric perturbation,which mainly modifies the gravity; on the other hand, $\theta$-dependent perturbations,  which is responsible for oblateness. Then all the equilibrium quantities, X, are well approximated by:
\begin{equation}
\label{dev} {\rm X} (\rm r,\theta)\, \simeq\,  \tilde{\rm X}(\rm r) + {\rm
X}_{22}(\rm r) \rm P_2(\cos \theta),
\end{equation}
The spherically symmetric part is then obtained by \citep[see for example][]{Kippenhahn1994}
\begin{equation}
\frac{\rm d\tilde{\rm p}}{\rm d \rm r}=-\tilde{\rho}\rm \, g_{\rm eff}, \hspace{0.3cm}\hbox{where }\hspace{0.3cm} \rm g_{eff}=\frac{GM_r}{r^2}-\frac{2}{3} \, \rm r \, \Omega^2  \nonumber
\end{equation}
The non-spherically symmetric part is obtained by (see DG92)
\begin{align}
\rm p_{22}&=-\tilde{\rho} \, r^2 \, \Omega^2 \, \left( \frac{\phi_{22}}{\rm r^2}\, \Omega^2+\frac{1}{3}\right) , \\
\rho_{22}&= \frac{\tilde{\rho}\, \rm \, r \, \Omega^2}{\tilde{\rm g}} \left( \frac{{\rm d} \ln \tilde{\rho}}{{\rm d} \ln \rm r}\right) \rm \left( \frac{\phi_{22}}{\rm r^2}\, \Omega^2+\frac{1}{3}\right), 
\end{align}
where
\begin{align}
\frac{1}{\rm r^2}\, \frac{\rm d}{\rm dr} \left( \rm r^2 \frac{\rm d \phi_{22}}{\rm d \rm r}\right) -\frac{6}{\rm r^2}\,\phi_{22} = 4 \, \pi \, \rm G \, \rho_{22}.
\end{align}
The boundary conditions can be found in \cite{Soufi1998} (hereafter S98).

\subsection{Oscillation frequencies up to cubic order in $\Omega$}

Using expansions of the type given in Eq.(\ref{dev}) for the oscillation
quantities, i.e., $\rm p'=\tilde{\rm p}' + \rm p_{2}'$, the
oscillation system is then expanded up to the cubic order. According
to S98 \citep[see also][]{K2008}, the oscillation equation then becomes:
\begin{align}
\label{Osc_eq}
\mathcal{L} \mathbf{\xi} = \left( \rm A+\epsilon \, \rm B\right) \mathbf{\xi} +\epsilon^2 \left(\rm D+\epsilon  \, \rm C \right) \mathbf{\xi} + O(\epsilon^4) = 0
\end{align}

\noindent $\epsilon$ being equal to $\Omega/\Omega_k$,
where $\Omega_k \, = \, \sqrt{\rm G \rm M / \rm R^3}$ is the break-up frequency.
The operator A represents the basic linear oscillation operator  including
the spherically symmetric perturbation due to rotation (through $\rm g_{eff}$).
The operators B and D, respectively, contain the effects of the Coriolis force and of non-spherically
symmetric distortion. The operator C shows that a coupling between the non-spherically symmetric distortion
and the Coriolis force exists. \\
Like in S98 and in \cite{K2008}, parts of the Coriolis and
centrifugal distortion effects are included into the
pseudo-zero$^{\rm th}$ order eigenvalue system. This way, we are
able to solve the eigenvalue problem up to cubic order without
having to solve the successive equations for the eigenfunctions at
each order. The solution yields eigenfrequencies $\sigma_0$, which
include parts of the frequency shifts induced by rotation. To single out various
contributions and emphasize the $m$-dependence, we write it in the form
\begin{align}
\sigma_{0,m} &= \sigma_0^{(0)} + \sigma_2^{(0)} + \sigma_{1,m} + \sigma_{2,m}^{\rm eigen} + \sigma_{3,m}^{\rm eigen}
\label{omega0}
\end{align}
where $\sigma_0^{(0)}$ is the classical zero$^{\rm th}$ order
frequency ignoring all effects of rotation,  $\sigma_2^{(0)}$ is the
correction resulting from the spherically-symmetric part of
the centrifugal distortion, and the next three terms give the contributions of
consecutive orders in $\Omega$ resulting from the Coriolis
acceleration. The linear term, $\sigma_{1,m}=m\,\Omega\,\beta\Omega_k$, is complete.
The higher order terms only include the parts resulting from the poloidal component of $\xi$.
The remaining contributions to frequency shifts
up to ${\cal O}(\Omega^3)$ are calculated as integrals involving the eigenvectors
$\xi_0$ (see S98).
To this accuracy, the complete expression for eigenfrequencies in uniformly
rotating star is given by
\begin{align}
\label{omega_tot}
\sigma_m &= \sigma_{0,m}+\sigma_{c,m} \\
\hbox{with    } \hspace{0.3cm} \sigma_{c,m}&= \sigma_{2,m}^{\rm T} +
\sigma_{2,m}^{\rm D} + \sigma_{3,m}^{\rm T} + \sigma_{3,m}^{\rm
D}+\sigma_{3,m}^{\rm C},
 \label{omegac}
\end{align}
where the exponent $^{\rm T}$ marks contributions from the Coriolis force acting on the
toroidal component of $\xi$, the exponent $^{\rm D}$ those arising
from the non-spherically symmetric distortion, and $^{\rm C}$ those resulting from coupling of the two
effects. \\

\subsection{Near degeneracy}
\label{Near_Deg_freq}

The standard perturbation approach is invalid if rotation couples
modes with close frequencies. Treatment of such cases requires
modification, which in the context of stellar pulsations was first used
by Chandrasekhar \& Lebovitz (1962) and developed later by DG92.
In the case of latitude-independent rotation profiles,
only  modes with the same $m$ and $\ell$s of the same parity are coupled.
In the present work, we study the coupling of two resonant modes
denoted  $k_j$ for $(n_j,\ell_j,m), \, j=1,2$ with frequencies $\sigma_1\ge\sigma_2$.
For the range of stellar models we are interested in, calculations
reveal that near degeneracy occurs for quite a large number of modes.

Near degeneracy is taken into account by searching for solutions of Eq.(\ref{Osc_eq}) in the form
\begin{align}
\vec{\xi} \, =& \, \sum_{k_j} \, \mathcal{A}_{k_j} \, \vec{\xi_{0,k_j}} \, + \, \vec{\xi_c} \hspace{0.7cm} j=1,2 \nonumber \\
\hbox{with} \hspace{1cm}& \\
\vec{\xi_c} \, =& \, \sum_{k \ne k_j} \, \alpha_{k} \, \vec{\xi_{0,k}} \,\hspace{1.5cm}  \, k=1,N, \nonumber
\end{align}
where the eigenfunction correction $\vec{\xi_c}$ is composed of all non-resonant modes.
The standard procedure leads to a linear system of equations  for $\mathcal{A}_{k_j}$ and the following
condition for a non-zero solution:

\begin{align}
\label{deg_det}
(\sigma_{k_1}\, - \, \sigma^{deg}_m) \, (\sigma_{k_2}\, - \, \sigma^{deg}_m)\, - \, \mathcal{H}_m^2 \, = \, 0
\end{align}
where the frequencies $\sigma_{k_1}$ and $\sigma_{k_2}$  are given
by Eq.(\ref{omega_tot}). The coupling term $\mathcal{H}_m$ corresponds to
 integrals containing second and third order contributions
(see S98 for more details). The solutions of Eq.(\ref{deg_det}), denoted by $\sigma^{deg}$,
  provide the desired eigenfrequencies:

\begin{align}
\label{ND_correction}
\sigma^{deg}_{m} \,&= \, \bar{\sigma}_m \, \pm \,  \sqrt{ \Delta_m^2 \, + \, \mathcal{H}_m^2 } \\
\hbox{where} \hspace{0.3cm} \bar{\sigma}_m \, &= \, \frac{\sigma_{k_1} \, + \, \sigma_{k_2}}{2} \hspace{0.3cm} \hbox{and} \hspace{0.3cm} \Delta_m \, = \, \frac{\sigma_{k_1} \, - \, \sigma_{k_2}}{2} \nonumber
\end{align}
where the sign $+$(resp.-) corresponds to mode $k_1$ (resp. $k_2$).
Note that if  $|\mathcal{H}_m|<<|\Delta_m|$, the
effects of coupling are small and mode frequencies are approximately described by Eq.(\ref{omega_tot}).

\section{Relative magnitude of the different contributions for uniform rotators}

In this section, we adopt as an example a stellar model with a simple equilibrium structure for which uniform rotation in depth is assumed.
All the results presented in this Section have been performed
for an $8.5 \rm M_{\odot}$ ZAMS stellar model described in Table \ref{tabZAMS}
rotating at $\Omega \simeq 15\% \, \Omega_k$ ($\rm v_{\Omega} \, = \, \rm R \, \Omega \, = \, 95 \, \rm km.s^{-1}$), which is representative for this type of star \citep{Stankov2005}.

\begin{table}[h]
\caption{Stellar parameters of the ZAMS model (Sect.2 and 3.)}
\label{tabZAMS}
\centering
\begin{tabular}{lccc}
\hline
$\rm M \, =  \, 8.5 \, \rm M_{\odot}$ &  $\rm R \, =  \, 3.96 \, \rm R_{\odot}$ \\
$\rm L \, =  \, 13 \times 10^{3} \, \rm L_{\odot}$ &  $\rm X_0 \, =  \, 0.7$ \\
$\rm P_c \, =  \, 8.8 \times 10^{3} \, \rm dyn.cm^{-2} $ &  $\rho_c \, =  \, 3.6 \times 10^{-9} \, \rm g.cm^{-3}$ \\
$\sigma_{\rm rot}\, = \, 15 \, \% \Omega_k$&  $\rm v_{\rm rot} \, = \, 95 {\rm km.s^{-1}}$\\
\hline
\end{tabular}
\end{table}

Stellar models and adiabatic oscillation frequencies are computed
with the evolution code CESAM2k \citep{Morel1997,Morel2008} and
the WarM (Warsaw Meudon) oscillation code, respectively (see S98). All numerical results
 presented here and in the rest of this paper concern  dipolar ($\ell=1$) modes.

\subsection{Frequency contributions up to cubic order}
\label{Sect_Cub_freq}
\begin{figure*}[th!]
\centering{
  \includegraphics[scale=1.2]{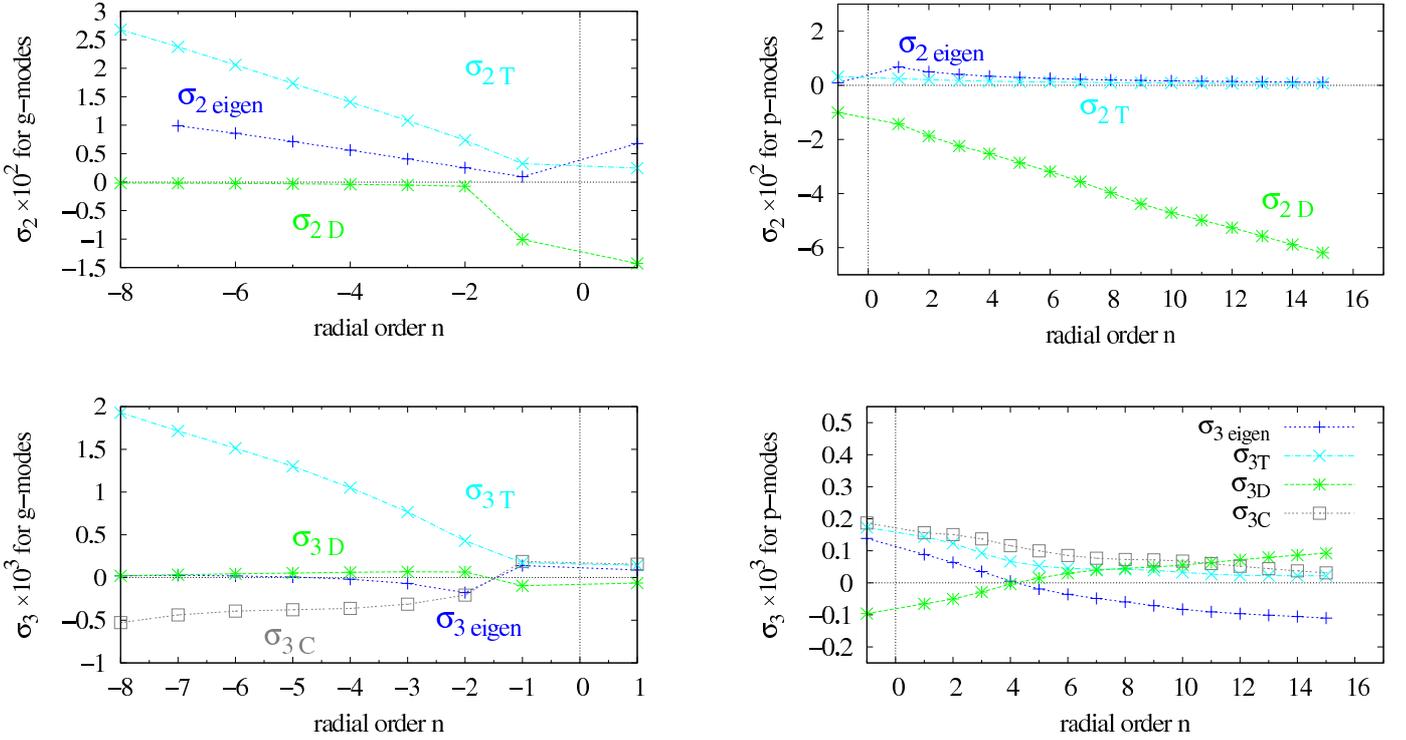}
  \caption{\label{Contrib_cub_2_sigma}
  Contributions of different approximation orders to $l=1,m=0$ mode frequencies:
   $\sigma_{2 \rm eigen}$ ($\sigma_{3 \rm eigen}$)
  represent the implicit $2^{\rm nd}$ ($3^{\rm rd}$ respectively)
  order contribution to the eigenfrequency;
  $\sigma_{\rm 2T}$ and $\sigma_{\rm 3T}$ denote $2^{\rm nd}$ and $3^{\rm rd}$ order frequency corrections due to the
   Coriolis acceleration;  $\sigma_{\rm 2D}$ and $\sigma_{\rm 3D}$
   represent $2^{\rm nd}$ and $3^{\rm rd}$ order frequency corrections due to the centrifugal distorsion;
    $\sigma_{3 \rm C}$ comes from the coupling of distorsion and Coriolis effects. The contributions are plotted
  as a function of the radial order n for an 8.5 M$_{\odot}$ ZAMS model rotating uniformly at $15 \% \, \Omega_k$,
  i.e. around 95 km s$^{-1}$ (see stellar parameters in Table \ref{tabZAMS}).
  The frequencies are scaled by $\Omega_k=\sqrt{\rm G \rm M/ \rm R^3}$: $\sigma = \omega / \Omega_{\rm k}$.
  We recall the use of a negative radial order n for gravity modes and a positive one for acoustic modes.
}}
\end{figure*}

Here we quantify the implicit contributions ($\sigma_{2,m}^{\rm eigen}$ and
$\sigma_{3,m}^{\rm eigen}$ from Eq.(\ref{omega0})), and
compare them with the corrective terms of the same orders given in Eq.(\ref{omegac}).
To extract these implicit contributions from the pseudo-zero$^{\rm th}$ order
eigenfrequencies we make use of their symmetry properties in $m$.
Thus, we get
\begin{equation}
\sigma_{2,m}^{\rm eigen} \, = \, \frac{(\sigma_{0, m}+\sigma_{0, -m})}{2} \, - \sigma_0^{(0)} \,
\end{equation}
and
\begin{equation}
\sigma_{3,m}^{\rm eigen} \, = \, \frac{(\sigma_{0, m}-\sigma_{0, -m})}{2} \, - \, m \frac{\Omega}{\Omega_k} \beta
\end{equation}
Figure \ref{Contrib_cub_2_sigma} shows all the contributions of different orders
to the rotational frequency corrections for g-modes ($\rm g_{14}$ to $\rm g_{1}$) and p-modes ($\rm p_{1}$ to $\rm p_{16}$).
We note the significantly different pattern for p- and g-modes.
In the latter case, the second order correction due
to the Coriolis acceleration dominates the rest. It grows linearly with the radial order
(in absolute value) as was already stated in \cite{Ballot2010} for polytropic models.
The Coriolis effect remains dominant at the third order, whereas distortion seems to have no effect.
The coupling between the Coriolis accelaration and distortion ($\sigma_{3C}$), 
is negative, thereby reducing the impact of the former.

For p-modes, the dominant second order term is $\sigma_2^{\rm D}$, as
 could be expected for modes which are mostly confined in superficial layers where the
the role of centrifugal acceleration is highest.
In this case too, the dominant term grows linearly with the radial order, as noted
 in \cite{Goupil2009} and \cite{Reese2006} for polytropic models.
 At the third order, $\sigma_{3}^{\rm eigen}$ and $\sigma_3^{\rm D}$ are of the same magnitude but with opposite signs
 and cancel  each other to some extent.

To sum up, as expected, for g-modes the most important contributions are related to the effects of the Coriolis acceleration, whereas
for p-modes, we must take into account both the implicit eigenfrequency terms (due to the part of Coriolis force included in the pseudo-zero$^{\rm th}$ eigen-system) and the effects of centrifugal distortion. In Table \ref{TabSig}, numerical values of
the different contributions are listed.

\subsection{Near degeneracy corrections}
Here we consider the same sequence of $\ell=1$ modes as in the previous section but
we now take into account the coupling of each mode with the nearest $\ell=3$ partner.
The coupled pairs must be of the same azimuthal order $m$.
 We use Eq.(\ref{ND_correction}) here to calculate the frequency shift
 caused by such a coupling.

\begin{figure}[t]
\centering{
  \includegraphics[angle=-90,scale=0.35]{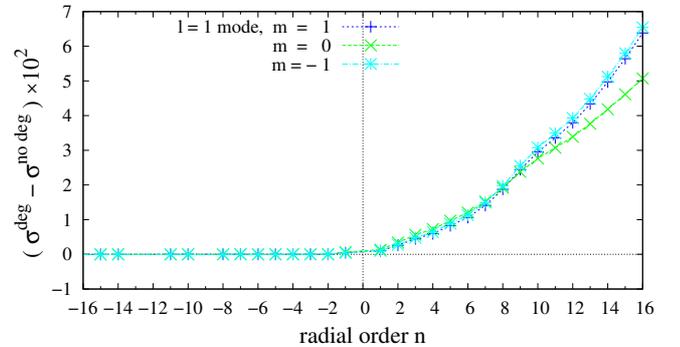}
  \caption{\label{Fig_deg_freq}Frequency differences between degenerate $\sigma^{\rm deg}$ and non-degenerate
  $\sigma^{\rm no \, deg}$ solutions scaled by $\Omega_k=\sqrt{\rm G \rm M/ \rm R^3}$
  ($\sigma = \omega / \Omega_{\rm k}$) as a function of the radial order, n, for
   $\ell=1$, $m=-1, 0, 1$ modes.
  The computations were made for the same model as in Fig. (\ref{Contrib_cub_2_sigma}).}}
\end{figure}

In order to compare the magnitude of this near degeneracy effect
with the second and third order contributions
shown in Fig.\ref{Contrib_cub_2_sigma},
 we depict in Fig \ref{Fig_deg_freq} the frequency differences
between  computations with and without near degeneracy being accounted for.
Figure \ref{Fig_deg_freq} shows that  near degeneracy primarily affects
the p-modes ($n > 0$).
This is not surprising as they are more sensitive to the outer regions which are more affected by 
distortion, a dominant factor in the coupling coefficient $\mathcal{H}$.
Moreover, this correction is found to be of the same magnitude
as the other second order corrections (see Fig.\ref{Contrib_cub_2_sigma})
but of the opposite sign. Hence the overall effect of the distortion is reduced.
However, this is not a universal property. As we may see in Eq.(\ref{ND_correction}),
the coupling always causes an increase of frequency separation between modes but the sign of
the shift is mode-dependent. In any case, rotational mode coupling is an important effect, especially for p-modes.
Taking it into account
\citep[as shown in][]{Suarez2010}  extends the validity domain of  perturbative methods.

Finally, we notice that the sectorial components of the $\ell=1$ triplet
are modified by roughly the same amount which implies that the rotational splitting
should not be strongly affected by near degeneracy (see Sect. \ref{s3_ss2}). This is expected
because in this case it enters as a third order effect.

\section{Rotational splitting for uniform rotators}

The rotational splitting can be defined as :  $S_{m}= (\sigma_{m}-\sigma_{0})/m$. 
 One also uses $\rm S_{m}= \sigma_{m}-\sigma_{m-1}$. 
 In this work, we use a scaled expression of the rotational splitting (Eq.(\ref{Sm})): 
\begin{equation}
\rm S_{m} = {\sigma_{m}-\sigma_{-m} / 2 m}
\label{Sm_scaled}
\end{equation}
These various definitions  are equivalent only at first  order in the rotation rate,
 $\Omega$, and equal to the linear splitting:
\begin{equation}
 S_{m} = \, \frac{1}{\Omega_k} \, \int_0^{\rm R} \int_0^\pi  K_{m}(\rm r,\theta) ~{\Omega(\rm r,\theta)} ~  \rm d\theta \rm dr
\label{sp}
\end{equation}
where the analytical expression for the kernels $\rm K_{m}$ is given in \cite{Goupil2010} and references therein. 
At higher orders in terms of $\Omega$, the two first definitions are contaminated by the 
effect of asphericity, which introduces an antisymmetric component in the frequency as a 
function of m. We choose to remove this second order contribution using the splitting expressed in Eq.(\ref{Sm_scaled}).

\begin{figure*}[th!]
\hspace{-1.5cm}  \includegraphics[angle=-90,scale=0.4]{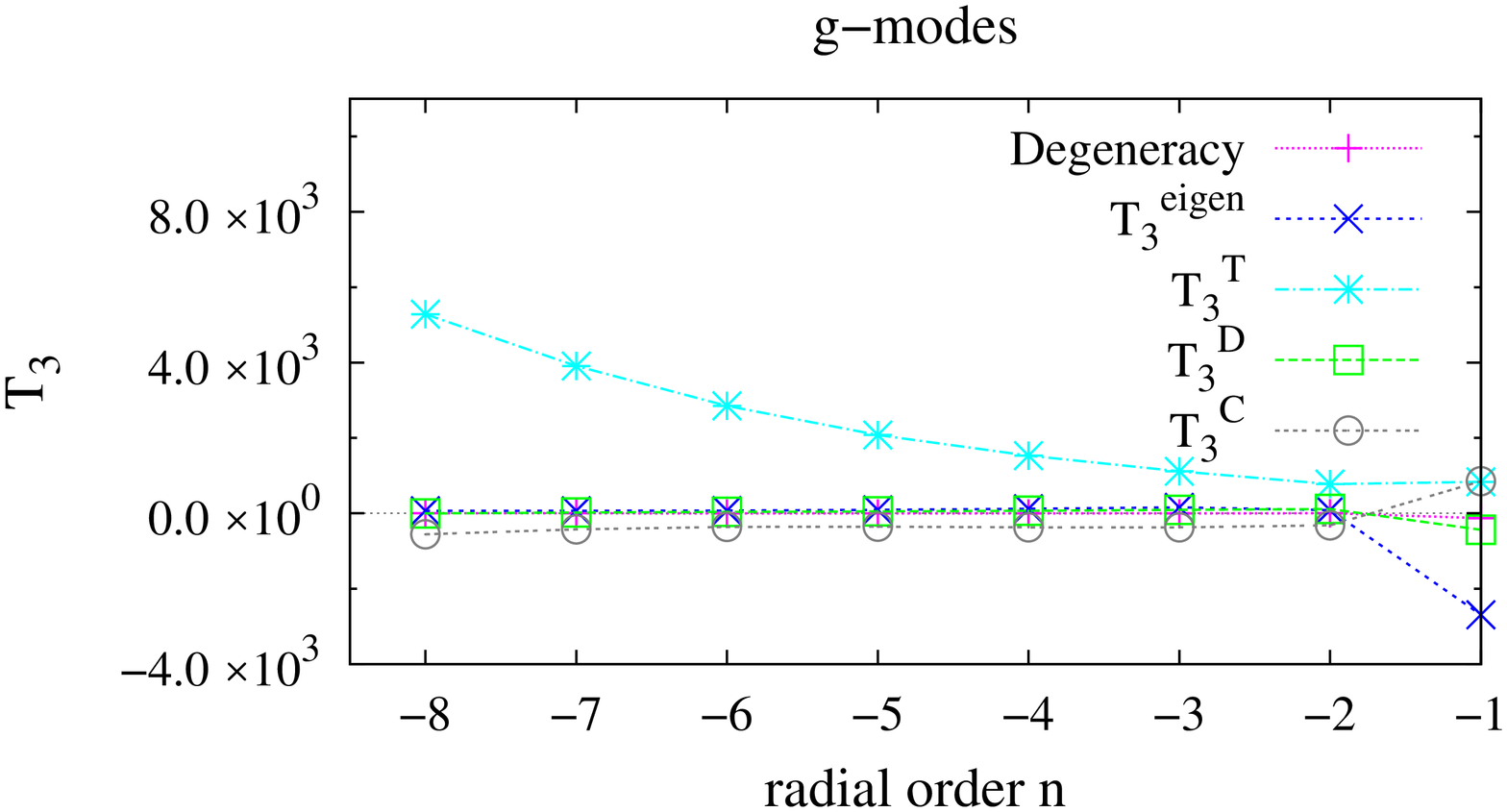}
  \includegraphics[angle=-90,scale=0.4]{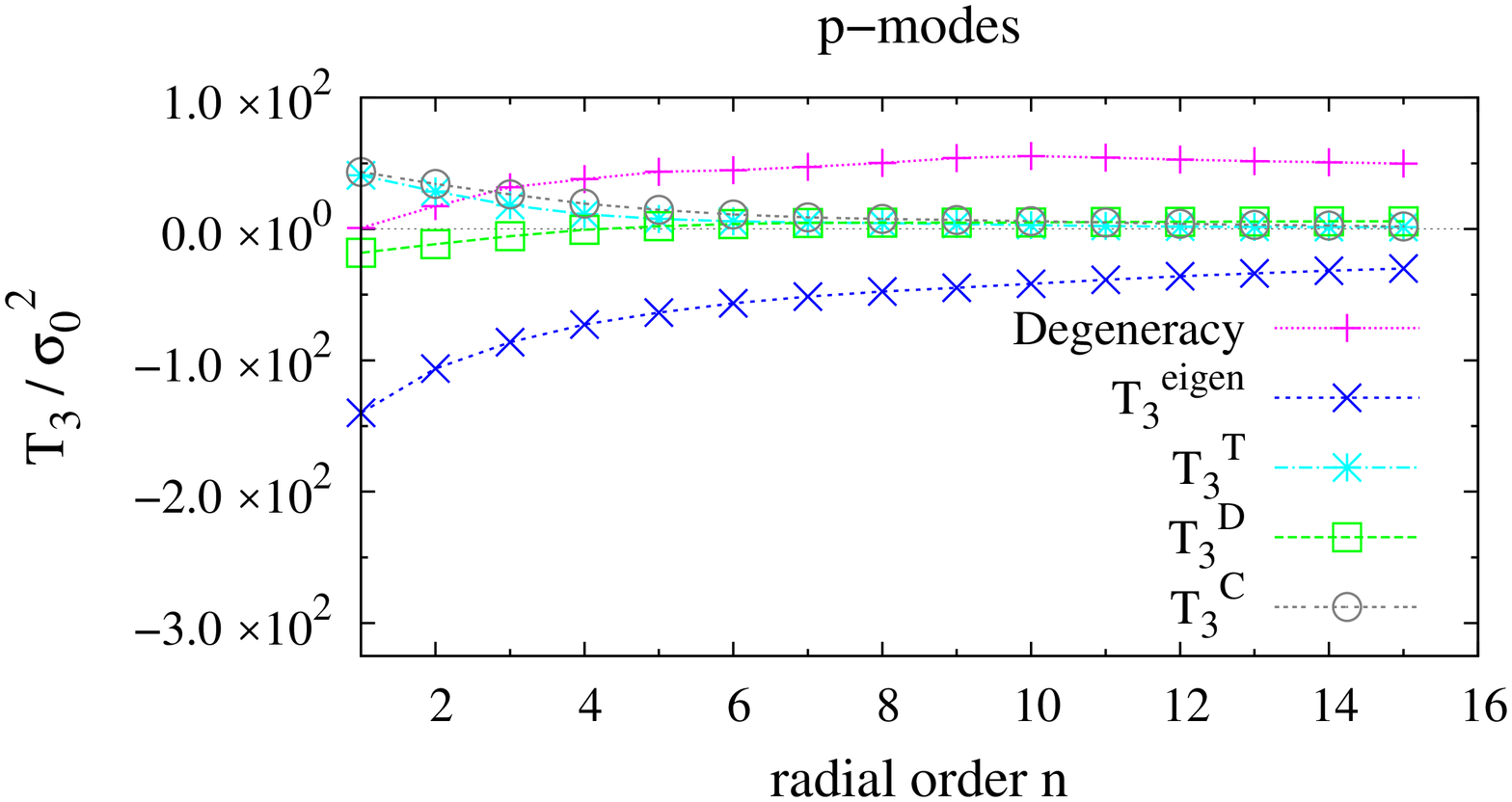}
  \vspace{-0.7cm}
  \caption{Scaled contributions to the splittings due to implicit third order terms
   in the eigenfrequencies $\left(\mathrm{T}_3^{\mathrm{eigen}}\right)$, the Coriolis effect $\left(\mathrm{T}_3^{\mathrm{T}}\right)$, distorsion $\left(\mathrm{T}_3^{\mathrm{D}}\right)$, the coupling of the two  $\left(\mathrm{T}_3^{\mathrm{C}}\right)$ and near degeneracy, \textbf{Left:} for $\ell=1$ g-modes, \textbf{Right:} for $\ell=1$ p-modes, divided by the square of the central  frequency. Computations were made for an 8.5 M$_{\odot}$ ZAMS model  uniformly rotating at $15 \% \, \Omega_k$, i.e.  $95 \, km.s^{-1}$ (see stellar parameters in Table \ref{tabZAMS}).}
\label{Figsplitcub} 
\end{figure*}

\subsection{Cubic order effects on the splitting}

Assuming a uniform rotation $\Omega=\Omega_0$, the splitting including frequency correction due to 
cubic order effects is given by:
\begin{align}
S_{m}^{\rm cubic} &=\frac{\Omega_0}{\Omega_k} ~ \beta +  
\frac{\Omega_0}{\sigma_0} \left(\frac{\Omega_0}{\Omega_k}\right)^2  \rm T_{3,m} \\
\rm T_{3,m}&=\rm T_{3,m}^{\rm eigen}+\rm T_{3,m}^{\rm T}+\rm T_{3,m}^{\rm D}+\rm T_{3,m}^{\rm C}
\label{Sm_cubic}
\end{align}
\noindent where $\sigma_0= \omega_0/\Omega_k$ is the normalized frequency of the corresponding axisymmetric 
 mode, $\Omega_0$ is the uniform rotation rate,
 and $\beta$ is the integral of the kernel $\rm K_{m}$ over 
 the star (see Appendix \ref{app_shear}, Eq.(\ref{def_beta})).

$\rm T_{3,m}$ contains the implicit third order contribution as well as effects due to the Coriolis acceleration, the distorsion, and the coupling of the two. 
From now on, we define  the departure from a linear splitting as follows: 
\begin{equation}
\label{dep_O3}
\delta \rm S_{m}^{\rm cubic} \equiv \rm S_{m}^{\rm cubic}-\left(\frac{\Omega_0}{\Omega_k}\right) \beta = 
 \frac{\Omega_0}{\sigma_0} \left(\frac{ \Omega_0}{\Omega_k}\right)^2 \rm T_{3,m}
\end{equation}

\subsection{Near degeneracy correction}
\label{s3_ss2}
According to the formalism explained in Section \ref{Near_Deg_freq}, 
if we consider the coupling of $\ell=1$ and $\ell=3$ modes, let the degenerate
 frequency of $\ell=1$ modes be:

\begin{align}
\sigma_{\ell=1,m}^{deg}\, &= \, \frac{\sigma_{\ell=1,m}+\sigma_{\ell=3,m}}{2} \, + \, \sqrt{\Delta_m^2 \,+ \, \mathcal{H}_m^2} \\
&= \, \sigma_{\ell=1,m} \, - \, \Delta_m \, + \, \sqrt{\Delta_m^2 \, + \, \mathcal{H}_m^2} \nonumber
\end{align}
where $\sigma_{\ell=1,m}$ and $ \sigma_{\ell=3,m}$ are non degenerate frequencies
given by Eq.(\ref{omega_tot}), and $\Delta_m$ is defined in Eq.(\ref{ND_correction}).
Then the splitting accounting for near degeneracy is given by:

\begin{align}
\label{NDsplit}
\rm S_{\ell=1,m=1}^{deg} \, =& \, \rm S_{\ell=1,m=1}^{ND} \, - \frac{1}{2}(\Delta_{1}\, - \, \Delta_{-1} )\\
&+ \, \frac{1}{2} \Bigl(\sqrt{\Delta_1^2 \,+ \, \mathcal{H}_1^2} \, - \, \sqrt{\Delta_{-1}^2 \,+ \, \mathcal{H}_{-1}^2}\Bigr)  \nonumber 
\end{align}
where
\begin{align}
 S_{\ell=1,m=1}^{ND} \, = \,  \frac{\sigma_{1,+1}-\sigma_{1,-1}}{2}
\end{align}
with $ND$ standing for non-degenerate. Note that $S_{\ell=1,m=\pm1}^{ND}$ contains cubic order contributions mentionned in the previous section. 
The contribution of near degeneracy to the splitting is then given 
by $\rm S_{\ell=1,m=1}^{deg} \, - \, \rm S_{\ell=1,m=1}^{ND}$. 

It is worth recalling  
that neglecting all cubic order contributions in Eqs. (\ref{omega_tot}), (\ref{omegac}) and 
(\ref{ND_correction})
results in $ \Delta_1=\Delta_{-1}$  and 
$\mathcal{H}_1=\mathcal{H}_ {-1}$. In that particular case, 
degeneracy does not contribute to the rotational splitting, 
and the rotational splitting is linear in $\Omega$ up to second order and satisfies :
\begin{align}
S_{m} &=\frac{\Omega_0}{\Omega_k} ~\beta 
\label{Sm_lin}
\end{align}

\subsection{Sensitivity to the nature of the eigenmode}

Figure \ref{Figsplitcub} displays the near degenerate contributions to the splitting for p- and g-modes 
together with the cubic ones for a ZAMS stellar model
(Table \ref{tabZAMS}).  The values of these different contributions 
are given mode by mode in Table \ref{TabT3_p}.  
  
  In Fig.\ref{Figsplitcub} (left),
  the Coriolis correction T$_{3}^T$ dominates for g-modes
  (by roughly a factor $10^2$ over other contributions) and decreases with 
  the radial order to a roughly constant value for low $|n|$. The 
  scale is too large in this figure to see the behavior
  of T$_3^{eigen}$, T$_3^D$, T$_3^C$ 
  and the near degeneracy contribution to the splitting,
  but we refer to Table \ref{TabT3_g} where it is shown that there 
  is no asymptotic behavior for these four contributions.
  Near degeneracy  is fully negligible  for all g-modes except for the $n=-1$
    mode that actually is a mixed mode.  
  The g-mode spectrum is much denser than the p-mode one but as shown in Fig. \ref{Hmplot}
  in Appendix \ref{App_HmDm}, the coupling term $H_m$ is much smaller than $\Delta_m$. This is due to
  the fact that distorsion effects are small for g-modes and therefore the overall
 (second  and third order) contribution to $H_m$ remains small.

In order to emphasize possible asymptotic behavior, 
the contributions to p-modes have been divided by $\sigma_0^{2}$. 
In Fig. \ref{Figsplitcub} (right), the implicit cubic order and the near-degenerate
contributions dominate for p-modes 
but with opposite signs and therefore roughly 
compensate each other. Hence, as was the case for pulsation frequencies,
the near degeneracy correction tends to reduce over-estimated contributions
to rotational splittings.
The contributions dominated by centrifugal 
distorsion, and in particular the near degeneracy one, scale as $\sigma_0^2$. 
In Appendix \ref{App_ND},  it is shown that  near degeneracy 
contributes to the rotational splitting  only if third order effects are taken into 
account (see Eq.(\ref{A_splitdeg_fin})). However it is also shown that 
second order  effects ($\mathcal{H}_2$) -- dominated by distorsion  ($\mathcal{H}_2$) 
for $p$ modes -- are involved. This explains why the near degeneracy 
frequency variation follows a $\sigma_0^2$ behavior for p-modes.

Similar conclusions for more evolved models with more complex structures are found for pure p-modes and pure g-modes. However these complex structures also give rise to mixed modes for which all effects contribute equally, and a precise investigation, mode by mode, has to be done for each  equilibrium model. This will be investigated in details in Sect. \ref{S4}.

\section{Effects of latitudinal shear on the splitting}

\begin{figure*}[th!]
\hspace{-1cm}  \includegraphics[angle=-90,scale=0.36]{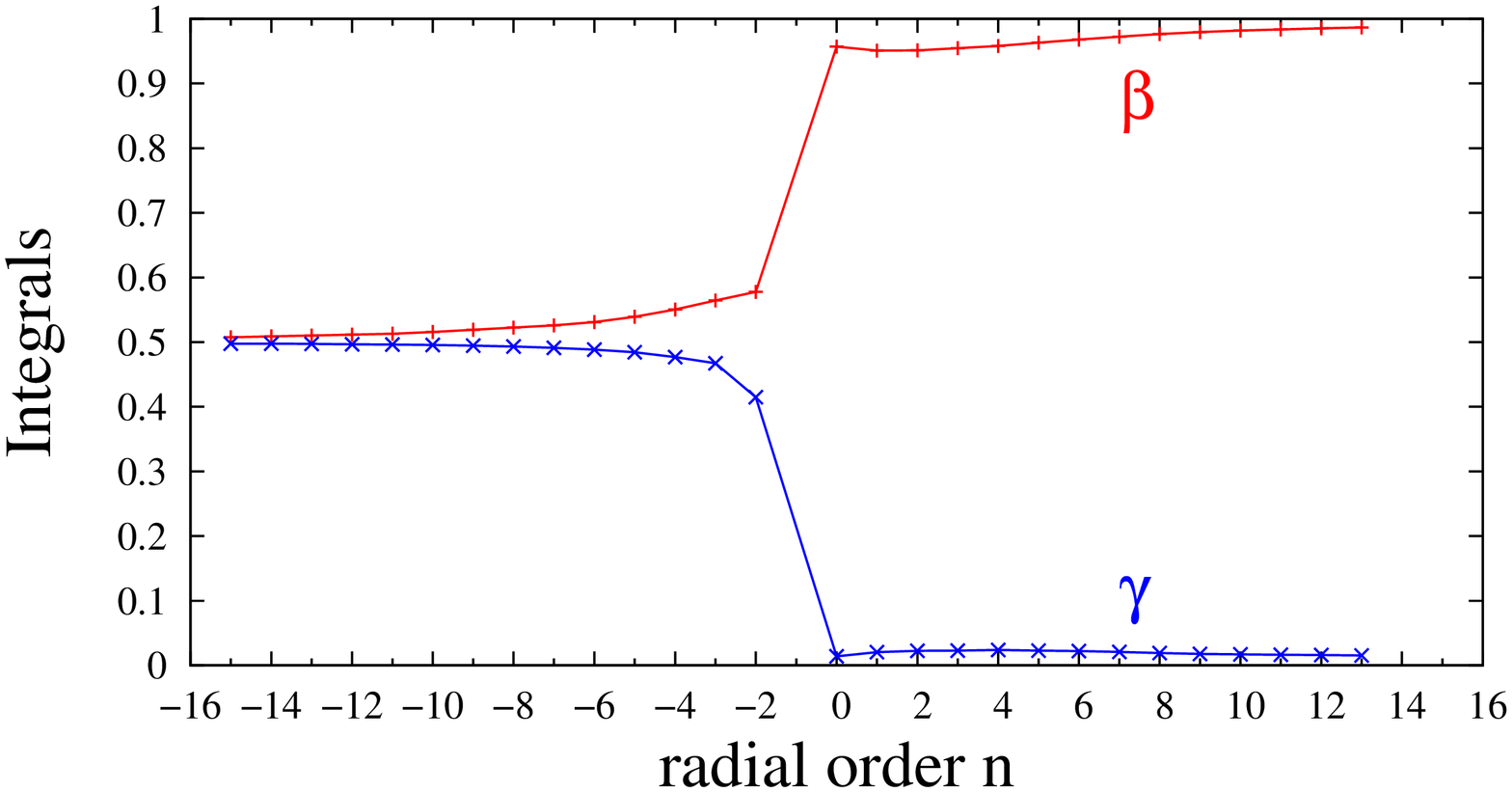}
  \hspace{1cm}
  \includegraphics[angle=-90,scale=0.36]{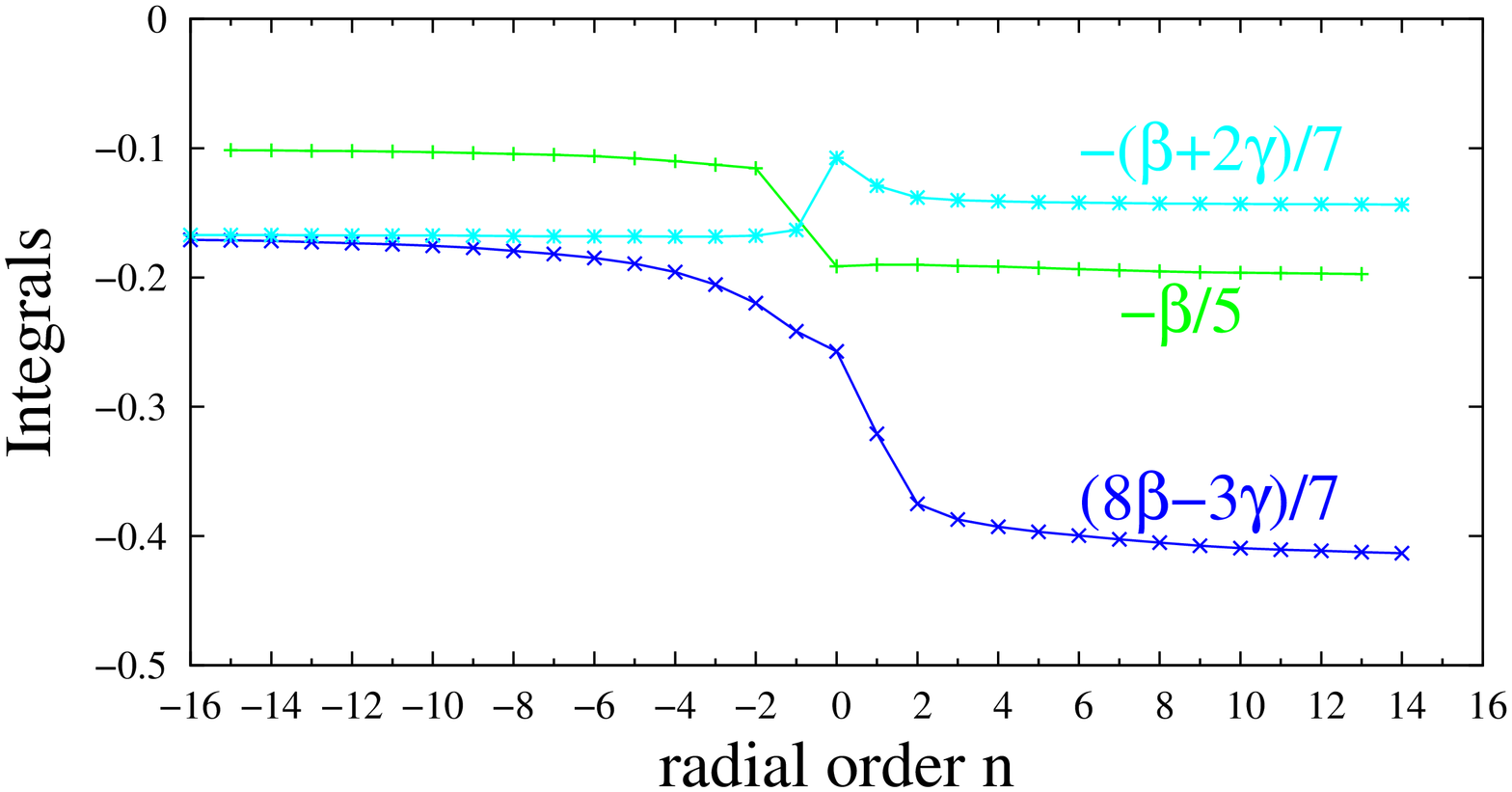}
  \vspace{-0.7cm}
  \caption{Behavior of \textbf{Left:} the scaled integrals $\beta$ and $\gamma$ (Eq.(\ref{eq_integ})) , \textbf{Right:} the scaled contributions to the splitting, as a function of the radial order n  (Eq.(\ref{S11}) to (\ref{S22})). Computations were made for an 8.5 M$_{\odot}$ ZAMS model uniformly rotating at $15 \% \, \Omega_k$, i.e. $95 \, km.s^{-1}$ (see stellar parameters in Table \ref{tabZAMS}).
  }
\label{Figbetgam} 
\end{figure*}

\cite{Hansen1977} derived the expression for  the rotational  splitting of adiabatic nonradial oscillations for slow  differential (steady, axially symmetric) rotation  $\Omega(r,\theta)$  and applied it  to numerical models of white dwarfs and  of massive main sequence stars   assuming a cylindrically symmetric rotation law.
  In the solar case, the  effects of latitudinal differential rotation on theoretical  frequencies  were investigated  by \cite{GT90},  \cite{DG91} and  \cite{DG92}.\\
In order to be able to compute the splittings from Eq.(\ref{sp}), one must specify a rotation law. It is convenient to assume the following form:
\begin{align}
\Omega(\rm r,\theta)&=  \sum_{\rm s=0}^{\rm s_{max}} ~ \Omega_{\rm 2s}(\rm r) ~(\cos\theta)^{\rm 2s}
\label{latit}
\end{align}
where $\theta$ is the colatitude.  The surface rotation rate at the equator is $\Omega(\rm r = \rm R,\theta=\pi/2) =\Omega_0(\rm r=\rm  R)$. \\
Note that in the solar case, $\Omega_2$ and $\Omega_4$ are negative and the equator rotates faster than the poles.\\
Inserting Eq.(\ref{latit}) into Eq.(\ref{sp}) yields the following expression for the splitting:
 \begin{align}
\rm S_{m}  &=  \frac{1}{\Omega_k} \, \int_0^{\rm R}   \Omega_0(\rm r) ~K(\rm r) ~\rm d \rm r+ \frac{1}{\Omega_k} \, \sum_{\rm s=0}^{\rm s=2}  m^{\rm 2s} ~\rm  H_{m,s}(\Omega) 
\label{Sm_gen}
 \end{align}
The expression for $\rm H_{m,s}(\Omega) $ can be found in Appendix \ref{app_shear}. \\

\ni
\subsection{Latitudinally differential rotation only $\Omega(\theta)$}
 In this case, for which rotation is assumed to be uniform in depth, the splitting becomes: \citep{Goupil2010}
\begin{align}
\rm S_{m}  &=   \frac{\Omega_{0}}{\Omega_k} ~ \beta  + \frac{1}{\Omega_k} ~\sum_{\rm s=0}^{\rm s=2}  m^{\rm 2s} ~  (\rm R_{\rm s}(\Omega) ~\beta + \rm Q_{\rm s}(\Omega)~ \gamma )
\label{split_rq}
\end{align}
Expressions for $\rm R_{\rm s}$ and $\rm Q_{\rm s}$ can be found in Appendix \ref{app_shear}. $\beta$ and $\gamma$ depend on the radial and horizontal components of the mode:
\begin{align}
\label{eq_integ}
\beta &= \frac{1}{\rm I} \int_0^{\rm R} \left[  \xi_{\rm r}^2-2 \xi_{\rm r} \xi_{\rm h} + (\Lambda -1) \xi_{\rm h}^2 \right]  \rho_0 \rm r^2 \rm d \rm r \\
\gamma&= - {1 / \rm I}\int_0^{\rm R} ~ \xi^2_{\rm h} ~ \rho_0 \rm r^2 ~ \rm d \rm r  
\end{align}
\ni
I being the inertia of the mode:
\begin{align}
\rm I &= \int_0^{\rm R} ~ (\xi_{\rm r}^2+\Lambda \xi^2_{\rm h}) ~ \rho_0 \rm r^2 ~ \rm d \rm r 
\label{inertia}
\end{align}
\indent
For the sake of simplicity, we restrict our study to $\rm s=1$ in Eq.(\ref{Sm_gen}). 
The rotation law can then be expressed as:
\begin{align}
  \Omega(\theta)=\Omega_0-\Delta \Omega \cos^2 \theta
\label{rotation_law}
\end{align}
Where $\Omega_0 = \Omega(\rm r = \rm R,\theta=\pi/2) $, $\Omega_2 = - \Delta \Omega$ and $\Omega_4=0$.
After some calculations provided in Appendix \ref{app_shear}, we are able to express 
the splittings of several $(\ell,m)$ combinations:
\begin{itemize}
\item[a.] $(\ell,m)=(1, \pm 1 )$ modes:
 \begin{align}
\rm S_{\ell=1,m= 1}^{\rm lat}  &=  \frac{\Omega_{0}}{\Omega_k} ~ \beta ~ (1 - \frac{1}{5}~ \frac{\Delta \Omega}{\Omega_{0}}) 
\label{S11}
 \end{align}
\item[b.] $(\ell,m)=(2, \pm 1 )$ modes:  
 \begin{align}
\rm S_{\ell=2,m= 1}^{\rm lat}  &=   \frac{\Omega_{0}}{\Omega_k} ~ \left( \beta ~ + \frac{1}{7} \frac{\Delta \Omega}{\Omega_0} ~ ( 8 ~ \gamma - 3 ~ \beta)\right) 
\label{S21}
 \end{align}
\item[c.]  $(\ell,m)=(2, \pm 2 )$ modes: 
 \begin{align}
\rm S_{\ell=2,m= 2}^{\rm lat}  &=   \frac{\Omega_{0}}{\Omega_k} ~ \left( 2\, \beta ~ - \frac{1}{7}  \frac{\Delta \Omega}{\Omega_0} ~ ( 2 ~ \gamma +~ \beta) \right)
\label{S22}
 \end{align}
\end{itemize}
In the solar case, $\beta \sim 1$ and  $\beta  >>|\gamma|$ 
  for the excited high frequency  p-modes.  Then the ($\ell=1,m=1$) splitting is:
\begin{align}
\rm S^{\rm lat}_{\ell=1, m=1}  &\approx  \,  \frac{\Omega_{0}}{\Omega_k} ~ \left[ 1+ \frac{1}{5} \left( {\Omega_2/ \Omega_{0}} +  {3/ 7} ~   {\Omega_{4}/\Omega_{0}}\right) \right] 
\label{S112}
 \end{align}
With $\Omega_2/\Omega_0=-0.127, \Omega_4/\Omega_0=-0.159$ , one obtains a departure
of $|{\rm S_1 / \Omega_0} -1|\simeq 0.04 $ from a linear splitting, i.e. a 4\% change in the solar case.

For upper main sequence stars, excited modes are around the fundamental radial mode  and may be 
mixed modes with $|\beta| \sim |\gamma| \sim 1/2$. This leads for instance to 
$|{\rm S_{1,1} / \Omega_0} -1|\sim 1 \% $  for $\Omega_2/\Omega_0 $ and $\Omega_4/\Omega_0$ equal to a half of the solar values.
We recall that for the stars treated in this article, we have taken $\Omega_2 = - \Delta \Omega$ and $\Omega_4=0$.\\

As shown in Eqs. (\ref{split_rq}) to Eq.(\ref{inertia}), 
the additional term due to the latitudinal shear strongly 
depends on the eigenfunction of the mode, through the radial and horizontal components of the displacement.
Therefore, we investigate the scaled contributions for different types of modes.
Again, the plots presented in Fig.\ref{Figbetgam} have been performed for an $8.5 \, \rm M_{\odot}$, $4 \, \rm R_{\odot}$ ZAMS model, rotating at $\Omega \simeq 15\% \, \Omega_k$ ($\rm v_{\Omega} \, = \, \rm R \, \Omega \, = \, 95 \, \rm km.s^{-1}$). Once again, computations for more evolved stellar models give similar results -- except for mixed modes around $\rm n=1$ -- even with stronger rotation rates. 

Figure \ref{Figbetgam} shows  the dependency of the $\beta$ and $\gamma$ integrals 
on the p or g nature of the mode. Globally, for g-modes, at a given rotation rate, we expect a small contribution of the 
latitudinal shear to the splittings, smaller for $S_{1,1}$ than for $\rm S_{2,1}$ and $\rm S_{2,2}$,
 whereas for p-modes, the contribution is quite important for $\rm S_{2,1}$.

For the sake of simplicity we will focus on $\ell=1$  splittings from now on. Note that the investigations presented in the 2 next Sections have been addressed for $\ell=2$ splittings as well and had lead to the same conclusions.

\subsection{A tachocline-like rotation profile: $\Omega(r,\theta)$}
\label{Tacho_theorie}
Let us refine the modeling of the rotation profile, assuming 
a rotation profile in depth as in the solar case. Rotation is uniform ($\Omega = \Omega_0$) in the inner layers and differential latitudinally- as expressed in (\ref{rotation_law})- in the convective envelope. $r_{cz}$ being the radius of the inner stable layer:
\begin{align}
\hbox{for} \hspace{0.3cm} r &< r_{cz}, \hspace{0.3cm} \Omega(r,\theta)\, = \, \Omega_0 \nonumber \\
r &\ge r_{cz}, \hspace{0.3cm} \Omega(r,\theta) \, =\, \Omega_0 \, - \, \Delta \Omega \, \cos^2 \theta
\end{align}
Therefore, after some calculations similar to those presented in Appendix \ref{app_shear}, equation (\ref{split_S11}) is no longer relevant, and should be replaced by:
\begin{align}
\label{split_lat_bz}
S_{\ell=1,m= 1}^{\rm lat}  \, &= \,\frac{\Omega_0 }{\Omega_k} \, \left( \beta \, - \, \frac{\Delta \Omega}{\Omega_0} \, \frac{1}{5} \, \beta_{cz}\right) \\
\hbox{where} \hspace{0.3cm} \beta_{cz}\, &= \, \frac{1}{\rm I} \int_{r_{cz}}^{\rm R} \left[  \xi_{\rm r}^2-2 \xi_{\rm r} \xi_{\rm h} + (\Lambda -1) \xi_{\rm h}^2 \right]  \rho_0 \rm r^2 \rm d \rm r 
\end{align}

Note that for p modes for instance, $\beta_{cz}$ is smaller than $\beta$ in the whole star -  $\beta_{cz}\sim 0.45$ whereas $\beta \sim 1$ -. Therefore in the case of a tachocline-like profile, the effect of a same latitudinal shear $\Delta \Omega$ is smaller than in the case where rotation is uniform in depth.

\section{The case of a $\beta$ Cephei on the main sequence}
\label{S4}
Massive stars on the main sequence are usually fast rotators and their fast rotation affects their internal structure 
as well as their evolution. Rotation of $\beta$ Cephei stars extends from  slow rotational velocity 
 ($\rm v<50$ km/s)   to extremely rapid ones ($\rm v> 250$ km/s) \citep{Stankov2005}.  
These stars usually have a radiative envelope  which may or may not be in latitudinal differential rotation. 
 For these fast rotators, one can wonder whether cubic order or 
near degeneracy contributions dominate over the effects of latitudinal shear. 
\begin{figure*}[t!]
  \centering{
    \includegraphics[scale=1]{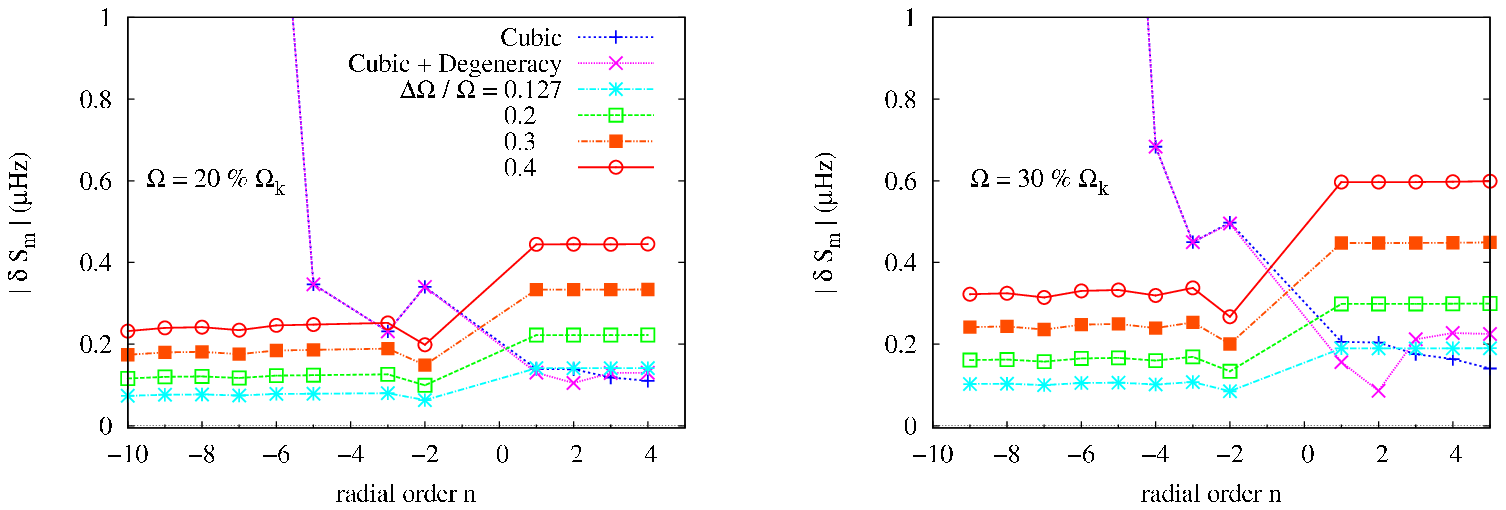}
    \caption[center]{\label{O3vslatfreqB} Departure from a linear splitting for ($\ell = 1, m = 1$) triplets, \textbf{Left}: for $\Omega\, = \, 20\% \, \Omega_k$, \textbf{Right}: for $\Omega\, = \, 30\%\, \Omega_k$, as a function of the radial order.  Different contributions result from: cubic order effects (dark blue),  cubic order effects with near degeneracy (purple, these two contributions overlap for all g-modes), and  latitudinally differential rotation with $\Delta\Omega/\Omega = 0.127\,$ (light blue), $0.2\,$ (in green), $0.3\,$ (in orange), $0.4\,$ (in red).  These results were computed for the uniformly rotating evolved 8.5 M$_{\odot}$ $\beta$ Cephei model described in Sect. 5.1 (Table \ref{tabBceph}).
    }
    \includegraphics[scale=1]{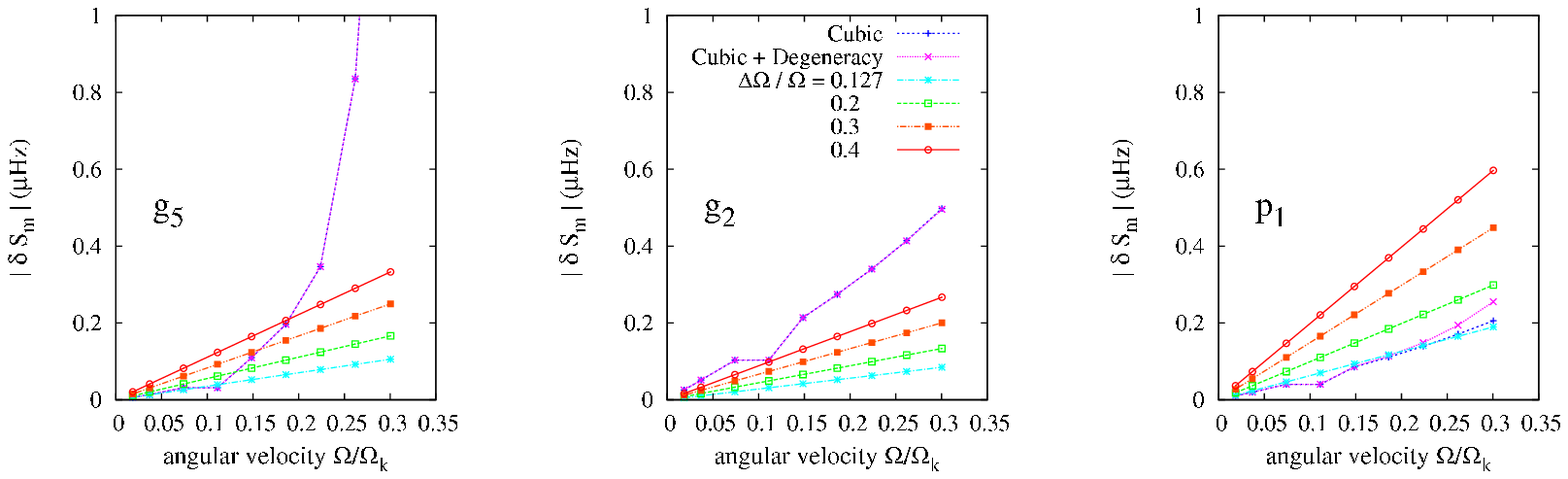}
    \caption[center]{\label{O3vslatrotB} Departure from a linear splitting for ($\ell = 1, m = 1$) triplets due to cubic order effects (dark blue), to cubic order effects including near degeneracy corrections (purple, these two contributions overlap for g$_5$ and g$_2$) and to latitudinally differential rotation with $\Delta\Omega / \Omega = 0.127\,$ (light blue) , $0.2\, $ (green), $0.3\, $ (orange), and $0.4\,$ (red).  These departures are plotted as a function of the rotation rate  for a $g_5$ mode (left), a $g_2$ mode (center) and a $p_1$ mode (right).  These results were computed for the uniformly rotating evolved 8.5 M$_{\odot}$ $\beta$ Cephei model described in Sect. 5.1 (Table \ref{tabBceph}). }}
\end{figure*}

Here we investigate the importance of  deviations from a linear splitting for
 an evolved main sequence model of an  $8.5\,\rm M_{\odot}$ star with a  $5.07 \, \rm R_{\odot}$
 radius (see Table \ref{tabBceph} for the stellar parameters of the model). 

\begin{table}[h]
\caption{Stellar parameters of the $\beta$ Cephei model in Sect.5}
\label{tabBceph}
\centering
\begin{tabular}{lccc}
\hline
$\rm M \, =  8.50 \, \rm M_{\odot}$ &  $\rm R \, =  \, 5.07\rm \, R_{\odot}$ \\
$\rm L \, =  2 \times 10^{37} \, {\rm erg.s^{-1}}$ &  ${\rm age} \, = \, 20 \, \rm My$\\
$\rm X_0 \, =  \, 0.713$ & $\rm Z_0\, = 0.014  $ \\
$\alpha \, =  \, 1.76 $ \\
\hline
\end{tabular}
\end{table}

\subsection{Departure from a linear splitting as a function of the frequency}
According to the region to which a mode belongs, very different types of behaviour are observed for the different contributions to the splitting, 
as illustrated in Fig.\ref{O3vslatfreqB}.

Figure \ref{O3vslatfreqB} displays the departure from a linear splitting due to cubic order effects with and without near degeneracy corrections, and to latitudinal differential rotation (uniform in depth). It has been computed for $\ell=1$ modes with radial orders ranging from -10 to 5. The parameters of the stellar model considered here are given Table \ref{tabBceph}.  It is rotating at $20 \% \Omega_k$ (left) and $30 \% \Omega_k$ (right). On these plots we observe three regions where the behavior of $\delta S_{\ell,m}$ differs. 

In the g-mode region ($n\le-2$), as the cubic order effect is proportional to  $\left(\frac{\Omega}{\Omega_k}\right)^2 \, \frac{\Omega}{\sigma_0}$ (see Eq.(\ref{dep_O3})), the smaller the frequency is, the higher cubic order terms are,  and this effect increases with the rotation rate. In this region, cubic order effects overtake those from latitudinally differential rotation.

In the p-mode region ($n>2$), latitudinal shear effects are larger than for g-modes. Cubic order effects are of same order of magnitude as the contribution from latitudinally differential rotation with a shear of only $12.7 \, \% $. 

In between, in the mixed mode region, cubic order effects with degeneracy corrections are of same order of magnitude as effects from latitudinally differential rotation for all the considered shears.

In order to get a better understanding of these different types of behavior, one can look at Fig.\ref{inertiaplot} in Appendix \ref{App_HmDm}. Figure \ref{inertiaplot} displays mode inertia (defined in Eq.(\ref{inertia})) as a function of the radial order, for the model described in Table  \ref{tabBceph}, uniformly rotating at 30$\, \% \, \Omega_k$. From this figure, the three frequency domains related to the nature of modes are clearly visible: below $g_3$ ($n \in \left[ -10,-3\right]$) are pure g-modes, above $p_2$ are pure p-modes, and in between is located the mixed mode region. 

Figure \ref{Hmplot}, which has been computed for the model described in Table  \ref{tabBceph}, uniformly rotating at 30$\, \% \, \Omega_k$, shows that near degeneracy has a very small effect on g-mode splittings: although frequency differences ($\Delta_m$) between $\ell=1$ and $\ell=3$ modes are small for g-modes (the y axis is in a logarithmic scale), the fact that $\mathcal{H}_m$ (which quantifies near degeneracy) is dominated by second order terms (due to distorsion) and takes small values for g-modes makes near degeneracy more or less negligible for g-modes. 
In the mixed mode region, although $\mathcal{H}_m$ increases,  the frequency differences between $\ell=1$ and $\ell=3$ modes $\Delta_m$ also increase. Consequently, near degeneracy should not be important in the range either. Once again, as mixed modes are very sensitive to the evolutionary stage, it is diffucult to make general statement for all the stellar models.  

In the p-mode domain, $\Delta_m$ decreases, while $\mathcal{H}_m$ increases causing the near degeneracy contribution to the splitting to increase. Therefore, the near degeneracy effect is larger in the pure p-mode region. This causes the departure from a linear splitting due to cubic order effects including near degeneracy  to be of the same order of magnitude as the latitudinal shear contribution (Fig. \ref{O3vslatfreqB}).

\subsection{Departure from a linear splitting as a function of the rotation rate}
We studied the relative values of third order, near degeneracy and latitudinal shear contributions
 to the splitting as a function of the rotation rate mode by mode.  
 Figure \ref{O3vslatrotB} shows three different cases: 
\begin{itemize}
\item the $\rm g_5$ mode illustrates the the case of high order g-modes (or pure g-modes) where the cubic order contribution overtakes the latitudinal shear contribution as soon as the mean rotation rate exceeds $15\, \% \, \Omega_k$ (which corresponds to a rotation velocity of around $80\, \rm km.s^1$).
\item the $\rm g_2$ mode is located in the mixed mode region where cubic order effects are larger than latitudinal differential rotation ones.
\item the $\rm p_1$ mode is still a mixed mode but with a nature closer to a pure p-mode for which near degeneracy is no longer negligible.  As a result, the total cubic contribution including degeneracy is of same order as a latitudinal shear of $12.7\, \% $.
\end{itemize}

\subsection{How do we disentangle the two effects?}
\begin{figure}[t!]
\centering{
  \includegraphics[scale=0.35,angle=-90]{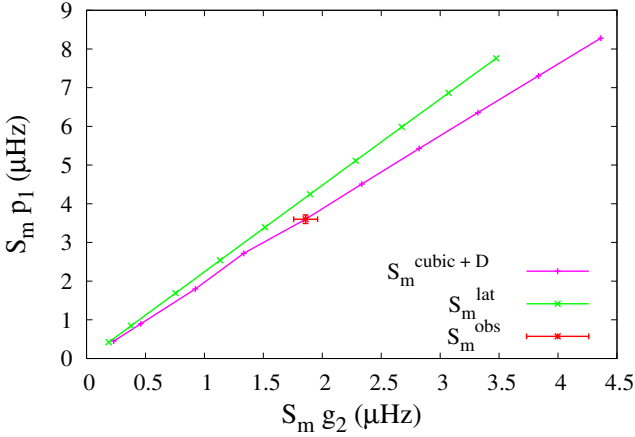}
  \caption[center]{ \label{distingoB}  $\rm S_{\ell,m}(p_1)$ versus $\rm S_{\ell,m}(g_2)$ due to cubic order effects and near degeneracy (purple), and to a latitudinal shear (green), for $\ell=1$ modes in an evolved 8.5 M$_{\odot}$ model (see Table \ref{tabBceph}).  The rotation rate increases from left to right: from 18$\% \Omega_k$ to 34$\% \Omega_k$.  The error bars for V1449 Aquilae have been taken from \cite{Degroote2009}.
  }}
\vspace{0.cm}
\end{figure}

In the previous subsection, we have shown that for a massive star on the main sequence, in the 
frequency range where we expect pulsation modes, it is not easy to conclude whether a departure
 from a linear splitting is due to latitudinally differential rotation or to cubic order 
 effects. Here we suggest a method to disentangle  the two effects.

Let us consider two modes: $g_2$ and $p_1$
 as representative of a mixed mode close to g-modes and a mixed mode close to p-modes, respectively, as seen in the previous section. 
  In Fig. \ref{distingoB}, $\rm S_{\rm 1}(\rm p_1)$ versus $\rm S_{\rm 1}(\rm g_2)$ is plotted for the 
two different assumptions (latitudinal shear or cubic order effects with near degeneracy corrections) 
for different rotation rates ($\Omega \le 35\% \Omega_k$).
Note that for g-modes, $\beta$ is roughly equal to $1/2$, and for p-modes $\beta$ approaches 1.
Accordingly,  the ratio between 
$\rm S_{\rm 1}^{\rm lat} (\rm p_1)$ and $\rm S_{\rm 1}^{\rm lat} (\rm g_2)$ does  
not depend on $\Delta \Omega$ and is approximatively equal to 2. On the other hand, the curve 
$\rm S_{\rm 1}(\rm p_1)$ as a function of  $\rm S_{\rm 1}(\rm g_2)$ for splittings including 
cubic order effects with near degeneracy corrections starts to deviate from a slope of 2 when the rotation rate is
large enough. The deviation increases with the rotation rate as expected. 

Let us now define $\rm S_1^{obs}$ as the observed splittings for  $l=1$ modes for 
a fast rotator  with a uniform rotation profile
$\Omega_0^{true}$. Their values are then assumed to be
given by  $\rm S_{\rm 1}^{\rm obs}=\rm S_{\rm 1}^{\rm D} \left( \Omega_0^{true} \right)$
  (Eq.(\ref{NDsplit})) i.e. rotational splittings computed up to cubic order,  including near degeneracy corrections.\\
If one misinterprets $\rm S_{\rm 1}^{\rm obs}$ as due to a latitudinal shear, 
$\rm S_{\rm 1}^{\rm obs}$ 
is assumed to obey  Eq.(\ref{S11}). The point 
representing $\rm S_m^{obs} (\rm p_1)$ as a function 
of $\rm S_m^{obs} (\rm g_2)$ ought to lie on the straigth line with slope 2
in Fig. \ref{distingoB}.  We represent this point  with the splittings 
$\rm S_m^{obs} (\rm p_1)$ and  $\rm S_m^{obs} (\rm g_2)$ computed according to 
Eq.(\ref{NDsplit}) assuming  a rotation rate of $15\% \Omega_k$, $v=80\, km.s^{-1}$. As the observed point is not located
on the straight line given by $\rm S_{\rm m}^{\rm lat} (\rm p_1) =  2 \rm S_{\rm m}^{\rm lat} (\rm g_2)$, 
we are then able to conclude that the deviation from a linear splitting is not due 
to a latitudinal shear. 

In the case of massive stars on the main sequence,
the effects of latitudinal differential rotation and cubic order or near degeneracy are 
of the same order of magnitude in the frequency domain where we expect to observe oscillation modes. 
This is mostly due to the mixed nature of modes around the fundamental frequency, 
and should therefore depend on the evolutionary stage of the star. But if two individual 
rotational splittings are available (one g-mode under the mixed mode region, and 
one p-mode above it), this method helps to distentangle whether a departure from a
linear splitting is due to cubic order contributions including accidental degeneracy, or latitudinal shear.

\section{The case of solar type stars}
\label{Sect_HD181}
\begin{figure*}[th!]
  \centering{
\hspace{-1cm}    \includegraphics[scale=1]{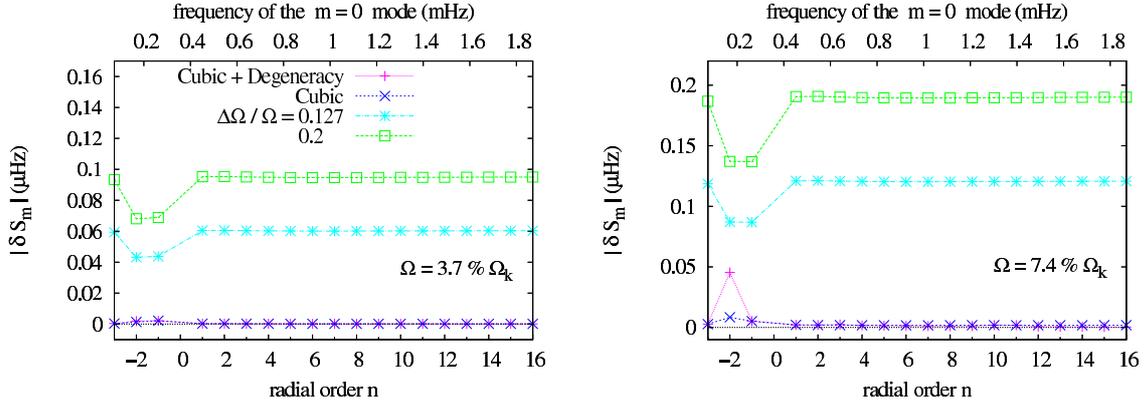}
    \caption[center]{\label{O3vslatfreqHD} 
    Departure from a linear splitting as a function of the radial order for ($\ell = 1, \rm m = 1$) triplets -- left: for $\Omega\, = \, 3.7\% \, \Omega_k$, right: $\Omega\, = \, 7.4 \%\, \Omega_k$ -- due to cubic order effects (blue), cubic order effects with near degeneracy (purble, the two of them overlap for most of the modes except g$_2$) and to a latitudinally differential rotation of $12.7\%$ (light blue) and $20\%$ (green). Computations were made for a model for HD 181420 ($M=1.36 \rm M_{\odot} $, see Table  \ref{tabHD181}).
    }}
  \vspace{0.3cm}
\end{figure*}

Low mass main sequence stars are known to be slow rotators. Indeed  due to their outer convection zone,  they undergo magnetic braking during their evolution \citep{Kawaler1988}. Observational evidence exists for  surface latitudinal differential rotation (coming from stellar spots due again to their outer convection zone). Hence, for these stars, the averaged surface rotation rate $\Omega$ is much smaller compared to that of more massive stars, such as $\beta$ Cephei stars, and $\Delta \Omega= \Omega_{equa}-\Omega_{pole}$, the difference between the rotation rates at the equator and the poles, can be large \citep[25\% for the Sun, between 1\% and 45\% for a star like  Procyon, ][]{Bonanno2007}. One therefore expects that latitudinal corrections to the splittings will dominate over cubic order effects which are negligible. 
As illustrative examples, we studied the case of HD 181906 and HD 181420, which are two solar like stars observed by CoRoT during the first long run, and which lightcurves have been analysed by \cite{Garcia2009} and \cite{Barban2009} respectively.

Before comparing the different contributions to the rotational splitting, one should wonder whether a perturbative approach up to the cubic order accounting for near degeneracy effects is valid for seismic interpretation purpose for this type of star. To answer this issue, we rely on the validity study done in \cite{Suarez2010}. This study has been performed for a polytropic model of $1.3$ M$_{\odot}$, that can be representative of HD 181420. Therefore we consider this study as well adapted in order to determine the validity of our approach for computing high order pressure pulsation modes which are propagating in the outer layers of a solar like star. In this work \citep{Suarez2010} show that for rotational velocities under approximately $40 \rm km.s^{-1}$, the structures of the frequency spectra computed with a non perturbative and their perturbative method are very similar. Here, we study the rotational splitting (that is a frequency difference which removes part of second order effects), with a cubic order perturbative approach accounting for the effect of near degeneracy. We then consider our approach as valid for rotation velocities at stake in the stars we study in this article.

\subsection{Competition between the three effects}

Here we investigate the order of magnitude of deviations from Eq.(\ref{Sm_lin}) for an $M=1.36\,\rm M_{\odot}$, $R=1.63 \, \rm R_{\odot}$ main sequence stellar model 
(see Table \ref{tabHD181} for the stellar parameters taken as a model representative of HD 181420). 

\begin{table}[h]
\caption{Stellar parameters of the model of HD181420 (Sect.6)}
\label{tabHD181} 
\centering
\begin{tabular}{lccc}
\hline
$\rm M \, =  \, 1.36 \, \rm M_{\odot}$ &  $\rm R \, =  \, 1.63 \, \rm R_{\odot}$ \\
$\rm L \, =  \, 4.4 \, \rm L_{\odot}$ &  $\rm X_0 \, =  \, 0.7$ \\
$\rm P_c \, =  \, 183 \rm \, dyn.cm^{-2} $ &  $\rho_c \, =  \, 5.1 \times 10^{-10} \, \rm g.cm^{-3}$ \\
\hline
\end{tabular}
\end{table}

In Fig. \ref{O3vslatfreqHD}, we plot the three kinds of contributions -- i.e. cubic order, cubic order with near degeneracy, and latitudinal shear -- 
to $(\ell=1, \pm 1)$ splittings as a function of the central mode frequency for two different rotational angular velocities ($3.7\% \, \Omega_k$ and $7.4\% \, \Omega_k$, which correspond to $15 {\rm km.s^{-1}}$ and  $30 {\rm km.s^{-1}}$, respectively, for the model described in  Table \ref{tabHD181}). Once again, these contributions show a peculiar behavior in the mixed mode region, where the two types of contributions can be of the same order. We focus here on higher frequencies, since the oscillation modes of HD 181420 are detected in the $1.5-2 \, \rm mHz$ frequency range \citep{Barban2009}.
Therefore, these plots show that even for rotation rates which are high for this type of star ($7.4 \% \, \Omega_k$), cubic order contributions (with or without near degeneracy) can be neglected compared to additional terms due to latitudinal shear. This leads to the conclusion that for HD 181420 in particular -- as we expect for solar-like stars in general -- the cubic order contributions to the splitting can safely be neglected in comparison with potential latitudinal shear contributions.
The same computations have been performed for a $1.2 \, \rm M_{\odot}$ $1.4 \, \rm R_{\odot}$ stellar model representative of HD 181906, and lead to the same conclusions.

\subsection{Determination of a latitudinal shear from seismic observations: }

The results of the seismic analysis provided in \cite{Barban2009} and \cite{Garcia2009} both find significantly different values for the low frequency peak in the Fourier spectrum and what is interpreted as the rotational splitting.
If a uniform surface rotation is assumed, it is not possible to reproduce such differences between these observables. 

\ni Let us assume a rotation profile of the form:
\begin{align}
\label{rotation_law2}
  \Omega(\theta)=\Omega_0 -\Delta \Omega \cos^2 \theta 
\end{align}
Where $\Omega_0$ is the rotation surface angular velocity at the equator, $\Omega_{equator}$, and $\Delta \Omega = \Omega_{equator}- \Omega_{pole}$. 
The rotation frequency is then given by :
\begin{align}
\label{Nu_rot}
  \nu_{\rm rot}(\theta)=\nu_{eq} \left( 1 -  \frac{\Delta \Omega}{\Omega_0} \, \cos^2 \theta \right)
\end{align}
where $\nu_{eq}$ and $\nu_{rot}$ correspond to the equatorial rotation rate
(i.e. $\Omega_0$) and the rotation frequency at the colatitude $\theta$ in $\mu$Hz respectively. 

\ni Using again the  rotation law Eq.(\ref{rotation_law2}) to derive the rotational splitting from Eq.(\ref{split_rq}), 
we obtain for $\ell=1, \rm m=\pm 1$:
\begin{align}
\label{split_S11}
\rm S_{1,1}^{lat}=\,\frac{\Omega_0}{\Omega_k} \, \beta \,\left(  1 - \frac{1}{5} \frac{\Delta \Omega}{\Omega_0} \right)
\end{align}
which can be written under the following form:
\begin{align}
\label{Nu_split}
\rm \nu_{\rm split}=\,\nu_{eq} \, \beta \,\left(\, 1\, - \,  \frac{\Delta \Omega}{\Omega_0} \, \right)
\end{align}
where $\nu_{split}$  corresponds to $\rm S_{1,1}$. 

\ni Then Eq.(\ref{Nu_rot})  and  Eq.(\ref{Nu_split})  are two equations 
with three unknowns: $\nu_{eq}$, $\cos \theta$ and $\Delta \Omega/\Omega_0$.
Dividing Eq.(\ref{Nu_split}) by Eq.(\ref{Nu_rot}) leads to : 
\begin{align}
\label{delta_Om_funct}
\frac{\Delta \Omega}{\Omega_0} = \frac{1\,-\,\beta d}{\cos{\theta}^2\, - \frac{1}{5} \, \beta \,d } 
\end{align}
Where we have introduced the parameter:
\begin{align}
\label{def_d}
d \, = \, \frac{\nu_{rot}}{\nu_{split}}
\end{align}
%As the latitudinal shear does not depend on the inclination angle $i$, it is not contaminated 
%by the uncertainties that are attached to the determination of $i$.
We then determine $\Delta \Omega / \Omega_0$ as a function of $\cos \theta$, 
along with the following constraints:
\begin{align}
\rm \theta \, \in \, \left[0;\frac{\pi}{2}\right] \, \Rightarrow \, \cos \theta \, \in \, \left[ 0;1 \right]\\
\frac{\Delta \Omega}{\Omega_0} \, \in \, \left[ -1;1\right]
\label{contraintes}
\end{align}

\begin{figure}[t!]
\includegraphics[angle=-90,scale=0.35]{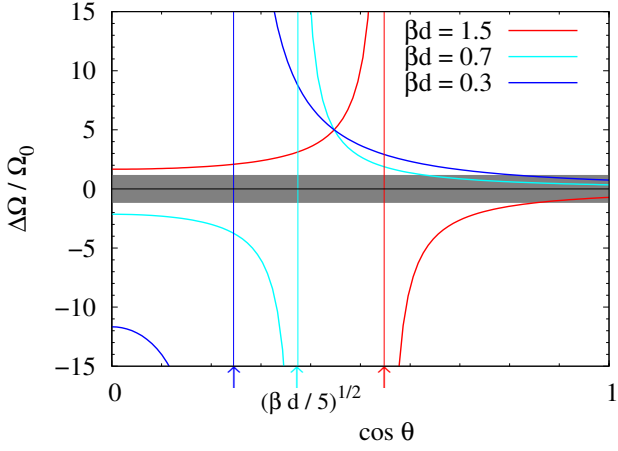}
\caption{Possible values for $\Delta \Omega / \Omega_0$, depending on the parameter d Eq.(\ref{def_d}) and the value of the integral $\beta$ Eq.(\ref{def_beta}). The line colors corresponds to the value of ($d \times \beta$) which ranges here from 0.3 (in blue) to 1.5 (in red). In grey is the acceptable domain of values for the latitudinal gradient (Eq. \ref{contraintes}). This plot has been performed for high order p modes.}
\label{lab_deltaOm} 
\end{figure}

As illustrated in Fig. \ref{lab_deltaOm}, the latitudinal shear is a hyperbolic function of $\cos \theta$, centered in $\sqrt{\beta d / 5}$. Depending of the values of the product $\beta d$ of the observationnal parameter $d$ and of the integral $\beta$, graphically we find different possible ranges of values for $\Delta \Omega / \Omega_0$ and $\cos \theta$.

\subsection{The case of HD 181906}

HD 181906 is an F8 dwarf for which fundamental parameters have been determined by \cite{Bruntt2009}, considering the presence of a background star : $m_{\rm v}=7.65$, $\rm L / \rm L_{\odot}=3.32\pm 0.45$, $T_{eff}=(6300\pm150) \rm K$, $\left[ \rm Fe / \rm H \right]=-0.11\pm 0.14 $, $\rm M/ \rm M_{\odot}=1.144 \pm 0.119$, $\rm v \sin \rm i=\left( 10 \pm 1 \right) \rm km.s^{-1}$. Note that \cite{Nordstrom2004} had found $\rm v \sin \rm i=\left( 16 \pm 1 \right) \rm km.s^{-1}$ considering HD 181906 as single, with no background star. It has been observed during the first long run of CoRoT, and  its light curve has been analysed by \cite{Garcia2009}. They find two possible interpretations for the mode identification, as listed in Table  \ref{tab_analysis_HD906}.

\begin{table}[h!]
\caption{Results concerning rotation from the analysis of HD 181906 performed in \cite{Garcia2009}, $\nu_{\rm rot1}$ and $\nu_{\rm rot2}$ stand for the two low frequency peaks attributed to the rotation of two different spots on the star surface. These values correspond to a product $\beta d$ ranging from $0.65\pm0.03$ to $0.70\pm0.03$.}             % title of Table
\label{tab_analysis_HD906} 
\centering                          % used for centering table
\begin{tabular}{c c c}        % centered columns (4 columns)
\hline
\hline                 % inserts double horizontal lines
~  & \hspace{-1.5cm}Scenario A & \hspace{-0.8cm}Scenario B \\
\hline                        % inserts single horizontal line
$\nu_{\rm rot1}\,(\mu \rm Hz)\,=$ & \hspace{2cm} $ 4.0 \, \pm \, 0.15$ \\
$\nu_{\rm rot2}\,(\mu \rm Hz)\,=$ & \hspace{2cm} $ 4.45 \, \pm \, 0.15$\\
$\nu_{\rm split} \,(\mu \rm Hz)\, =$ & \hspace{-1.5cm}$ 5.8 \, \pm \, 0.14 $ & \hspace{-0.8cm}$6.1 \, \pm \, 0.14$ \\
\hline                                   %inserts single line
\end{tabular}
\end{table}

Considering the uncertainties attached to the seismic observables, the result obtained for the latitudinal shear is presented in Tab. \ref{tab_res_HD906}. These results show that the hypothesis of latitudinal differential rotation is consistent with the available observables . Therefore, the latitudinal differential rotation profile is a reliable assumption concerning HD 181906. Moreover, The results define a range of possible values for different characteristics of the rotation profile: the latitudinal shear, the rotation surface velocity at the equator, as well the inclination angle. The correspondant error bars are discussed in Sect. \ref{S_Discussion}. Both scenarii give the same conclusionsvery similar results.
\begin{table}[h!]
\caption{Latitudinal shear, equatorial velocity and inclination angle obtained for HD 181906, using models with rotation uniform in depth. The two last lines correspond to $i = \arcsin \left(v \sin i / v_{eq} \right)$ with $v \sin i\, =\, (10 \pm 1)\, {\rm km.s^{-1}}$ \citep{Bruntt2009}, or with $v \sin i\, =\, (16 \pm 1)\, {\rm km.s^{-1}}$ \citep{Nordstrom2004}. Note that the two spots give hardly the same results, we present here mean values computed for the two surface rotation velocities. }             % title of Table
\label{tab_res_HD906} 
\centering                          % used for centering table
\begin{tabular}{c c c}        % centered columns (4 columns)
\hline
\hline                 % inserts double horizontal lines
~  & Scenario A & Scenario B \\
\hline                        % inserts single horizontal line
$\Delta \Omega / \Omega_0 \, =$ &  $ 0.55 \, \pm \, 0.21 $ & $0.59 \, \pm \, 0.18$ \\
v$_{eq}\, ({\rm km.s^{-1}})\, =$ & $ 41.3 \, \pm \, 2.5 $ & $43.8 \, \pm \, 1.6$ \\
$i \hspace{0.3cm}(\hbox{Brunt 2009}) \, (^{\rm o})\, =$ & $26 \, \pm \, 2$ & $24 \, \pm \, 2$ \\
$i \hspace{0.3cm}(\hbox{Nordstrom 2004})\, (^{\rm o})\, =$ & $23 \, \pm \, 3$ & $21 \, \pm \, 2$ \\
\hline                                   %inserts single line
\end{tabular}
\end{table}

\subsection{Solar or anti-solar latitudinal shear for HD 181420 ?}

HD 181420 is an F2 main sequence star whose fundamental parameters have been determined by \cite{Bruntt2009} : $m_{\rm v}=6.57$, $\rm L / \rm L_{\odot}=4.28\pm 0.28$, $T_{eff}=(6580\pm105) \rm K$, $\left[ \rm Fe / \rm H \right]=0.00\pm 0.06 $, $\rm M/ \rm M_{\odot}=1.31 \pm 0.063$, $\rm v \sin \rm i=\left( 18 \pm 1 \right) \rm km.s^{-1}$.
It has been observed during the first long run of CoRoT, and  its light curve 
has been analysed by \cite{Barban2009}. They find two possible interpretations for the mode identification, as listed in Table  \ref{tab_analysis_HD}.  Later Bedding \& Kjeldsen (2010) used some empirical scaling method that seemed to favor scenario 1. 

\begin{table}[h!]
\caption{Results concerning rotation from the analysis of HD 181420 performed in \cite{Barban2009}.}             % title of Table
\label{tab_analysis_HD} 
\centering                          % used for centering table
\begin{tabular}{c c c}        % centered columns (4 columns)
\hline
\hline                 % inserts double horizontal lines
~  & \hspace{-1.5cm}Scenario 1 & \hspace{-0.8cm}Scenario 2 \\
\hline                        % inserts single horizontal line
$\nu_{\rm rot}\,( \mu \rm Hz)\,=$ & \hspace{2cm} $ 4.5 \, \pm \, 1.5$\\
$\nu_{\rm split} \,( \mu \rm Hz)\, =$ & \hspace{-1.5cm}$ 2.59 \, \pm \, 0.38$ & \hspace{-0.8cm}$ 3.29 \, \pm \, 0.17$ \\
\hline                                   %inserts single line
\end{tabular}
\end{table}

Concerning HD 181420, our model gives unexpected results: in both scenarii, the latitudinal shear is found negative, i.e. the pole rotate faster than the equator, which is the opposite of what is known for the Sun. In order to appreciate the reliability of this result, one should refer to the work of \cite{Kapyla2011}, who study the impact of rotation on turbulent angular momentum and heat transport in solar like stars convective zone,  by the mean of direct numerical simulations of turbulent convection in spherical geometry. According to the authors, the rotation profile varies from anti-solar (equator rotates slower than poles) for low Coriolis number to solar (equator rotates faster), with a transition around $\rm Co = 3$. The Coriolis number which measures the impact of rotation on turbulent motion is given by:
\begin{align}
{\rm Co}\, = \, 2 \, \Omega_0\, \tau_c
\end{align} 
Where $\tau_c$ is the convective overturning time scale -computed in the stellar evolution code CESAM \citep{Morel2008}-, and $\Omega_0$ is taken as the angular velocity at the equator. For HD 181420, when rotation is supposed to be uniform in depth, the Coriolis number is $(3.2\, \pm 0.5)$ for scenario 1 and $(4.0\, \pm 0.3)$ , i.e. both values are only slightly above the transition threshold between anti-solar and solar surface rotation \citep[see Fig. 17 of ][]{Kapyla2011}. Therefore, it is difficult to confirm that the physics at stake in the convective envelope of HD 181420 can lead to anti-solar rotation profile for HD 181420.

Let us refine the modeling of HD 181420, taking a tachocline-like profile. As explained in Sect. \ref{Tacho_theorie}, it consists in assuming uniform rotation ($\Omega=\Omega_0$) in the inner layers, and latitudinal differential rotation ($\Omega(\theta)$) in the convective envelop. The rotational splitting given by Eq.(\ref{split_S11}) is no longer relevant, and should be replaced by Eq. (\ref{split_lat_bz}).

In this case, we found different values for $\Omega_0$ than in the former case, as a consequence, the Coriolis number reaches $(8.9\, \pm 1.7)$ for scenario 1 and $(11.8\, \pm 0.7)$ for scenario 2, i.e. high above the transition limit between anti-solar and solar rotational shear. The integral $\beta_{cz}$ reaches a constant value of 0.45 for high order p-modes, which changes the relationship between the colatitude of the spot and the latitudinal shear (\ref{delta_Om_funct}), and as a consequence changes the domain of possible shear.
Positive latitudinal shear corresponding to solar type latitudinal rotation are obtained for the two scenarii, which is this time fully consistent with the Coriolis number values. The results obtained with this tachocline-like model are listed in Tab. \ref{tab_res_HD420}. 

\begin{table}[h!]
\caption{Latitudinal shear, equatorial velocity and inclination angle obtained for HD 181420, using models with tachocline-like rotation profile. The last line corresponds to $i = \arcsin \left(v \sin i / v_{eq} \right)$ with $v \sin i\, =\, (18 \pm 1)\, {\rm km.s^{-1}}$ \citep{Bruntt2009}}             % title of Table
\label{tab_res_HD420} 
\centering                          % used for centering table
\begin{tabular}{c c c}        % centered columns (4 columns)
\hline
\hline                 % inserts double horizontal lines
~  & Scenario 1 & Scenario 2 \\
\hline                        % inserts single horizontal line
$\Delta \Omega / \Omega_0 \, =$ &  $ 0.50 \, \pm \, 0.45$ & $0.66 \, \pm \, 0.11$ \\
$\nu_{eq}\,(\mu \rm Hz)\, =$ & $1.0 \, \pm \, 0.2$ & $1.3 \, \pm \, 0.1$ \\
v$_{eq}\, ({\rm km.s^{-1}}) =$ & $45.6 \, \pm \, 8.7$ & $60.1 \, \pm \, 3.5$ \\
$i \hspace{0.3cm}(\hbox{Brunt 2009}) \, (^{\rm o})\, =$ & $23.2 \, \pm \, 7.4$ & $17.4 \, \pm \, 2.0$ \\
\hline                                   %inserts single line
\end{tabular}
\end{table}

The case of HD 181420 is particularly interesting as a simple model of rotation uniform in depth and differential in latitude leads to non-physical latitudinal shear. Only a little more sophisticated model, where rotation varies in depth according to a tachocline-like profile, is fully coherent with the physics of the convective zone as well as the observables. We also give a range of possible values for the rotation profile parameters: the latitudinal shear in the convective zone, the rotation surface velocity at the equator and the incination angle. To conclude, for this star, we find that rotation profile inside the star should rather be a tachocline like profile, with a solar type latitudinal rotation. Moreover, these results, when compared to those of \cite{Mosser2009}, seem to favor scenario 1 since the rotational frequency at the equator in the hypothesis of scenario 1 is closest to the one found in \cite{Mosser2009} by spot modelling - $ \left( 5.14 \, \pm \, 0.07 \right) \times 10^{-6} \, {\rm rad.s^{-1}}$-. 

\subsection{Discussion}
\label{S_Discussion}
As already mentionned in the beginning of this section, \cite{Suarez2010} have found that second order perturbative methods including near-degenaracy corrections are valid for rotation velocity up to approximately $40 \rm km.s^{-1}$. In this study, third order perturbative methods have been used for the computation of rotational splittings in Sect. 7.1. In Sect. 7.2, is given a simple recipe which allows to compute the latitudinal shear given seismic observables. This formulation rely on the validity of the third order perturbative method, but the only quantity provided by the modeling are $\beta$ (Eq. \ref{def_beta}) and $\beta_{cz}$, i.e. quantities computed from first order eigenfunctions. The question is then: are eigenfunctions suffitiently varing with rotation rate (under $60 \rm km.s^{-1}$ i.e. 15 \% $\Omega_k$), to impact the value of $\beta$?  After verifications, it appears that for high order acoustic modes in moderately rotating stars (under 15$\% \Omega_k$) $\beta \simeq 1$, and $\beta_{cz} \simeq 0.45$ are satisfying evaluations. Therefore the recipe is valid for rotation rates at stake in Sect. 7.3 and 7.4. Note the convenience of the proposed recipe which only need for seismic observables, and the values of $\beta$.

The large error bars obtained in particular for the values of latitudinal shears (Tab. \ref{tab_res_HD906}, \ref{tab_res_HD420}) is not only due to the observational uncertainties, but it can also be attributed to the simplicity of the spot model used.
This model does not account for spot parameters such as for example a spot lifetime, or the spot distribution on the observed stellar disc \citep[for more sophisticated modelling see][]{Mosser2009}.  We only consider a unique spot of infinite lifetime.
Note that the use of a more complicated model would require more observational constraints than only one spot rotation signature. When only mean values of rotationnal splittings are available, and the observational error bars on the low frequency signature of a spot rotation are large \citep{Barban2009}, we are able to give general conclusions concerning the rotation profile -i.e. uniform in depth or tachocline like, solar or anti-solar-, but no reliable numerical values caracterising it.

\section{Conclusions}
\label{S_CCL}
With the help of the perturbative approach established in \cite{Soufi1998}, we investigated
 second and third order contributions of the Coriolis and 
the centrifugal accelerations to p and g-modes frequencies, as well as near degeneracy effects 
on the rotational splittings. Their effects were then compared with 
those of a latitudinal shear. 
We studied two types of oscillating stars. 

\indent For an evolved model of a $\beta$ Cephei, the effects of near degeneracy, cubic order perturbations
and  latitudinal shear are of the same order of magnitude in the frequency range, relevant to
such stars -- i.e. low order p- and g-modes -- and for reasonable rotation
rates ranging from $15\%\, \Omega_k$ to $30\% \, \Omega_k$. Nevertheless, when two individual
splittings for  modes of a different nature (a pure g and a mixed mode for example) 
are available, a method is proposed to distinguish between a latitudinal shear and the other effects. 

\indent
For solar-like stars such as HD 181420 and HD 181906, which are mostly moderate rotators and oscillate with high order p 
modes, cubic order effects on frequency splittings are shown to be small in front of the effects of latitudinally
differential rotation. Therefore, given a splitting and a rotation period signature, it is possible 
to infer a range for the latitudinal shear coefficient $\Delta \Omega / \Omega_0$. 
Although no precise determination of latitudinal shear is possible unless the spot latitude
is fully determined, we have been able to determine the most reliable rotation profile for each of the two stars. More over the determination of the latitudinal shear by our seismic method can be taken as a constraint to be added to other observational constraints, such as those coming from spectropolarimetric results  \citep[e.g.][]{Donati2010} or spot modelling  \citep[e.g.][]{Mosser2009}.

\begin{acknowledgements}
We first would like to thank the referee for relevant remarks which helped to improve this article. We are indebted to W.D. Dziembowski for his helpful remarks, and to D. Reese for his careful reading of the manuscript. We also thank St\'ephane Mathis for helpful discussions about transport of angular momentum in solar like stars convective envelope.
\end{acknowledgements}

\begin{appendix}

\section{Near degeneracy corrections to the splittings of high order p-modes}
\label{App_ND}
This section is dedicated to a qualitative estimation of the near degeneracy corrections to 
rotational splittings:
\begin{align}
\label{A_splitND_init}
\rm S_{\ell=1,m=1}^{deg} \, - \, \rm S_{\ell=1,m=1}^{ND} \, & =
 \, \frac{ \sqrt{\Delta_1^2 \,+ \, \mathcal{H}_1^2} \, - \, \sqrt{\Delta_{-1}^2 \,+
  \, \mathcal{H}_{-1}^2}}{2}  \, \nonumber \\
& -\frac{1}{2} (\Delta_{1}\, - \, \Delta_{-1})
\end{align}
where $\Delta_1$ and $\mathcal{H}_1$ are normalized by the break-up frequency $\Omega_k$.
The oscillation frequency given in Eq.(\ref{omega0}) can be rewritten in the following the generic form :
\begin{align}
\sigma_{n,\ell,m}\, = \, \sigma_{0,n,\ell}^{0} \, &+ \, m \frac{\Omega}{\Omega_k} 
\beta_{n,\ell}\, + \, m^2 \,  d_{n,\ell,|m|}\,   + \, m
 \,  t_{n,\ell,|m|}\, \nonumber
\end{align}
where $d_{n,\ell,|m|}$ and $t_{n,\ell,|m|}$  represent the second and third order contributions.
Therefore the term $\Delta_1$  and $\Delta_{-1}$ appearing in Eq.(\ref{NDsplit}), for the coupling of 
an $\ell=1,n$ and an $\ell=3,n'$ mode, can be written as:
\begin{align}
\Delta_1 \, = \, \Delta^0 \, &+ \,   \frac{\Omega}{\Omega_k} (\beta_{n,1}-\beta_{n',3})\,  
+   (d_{n,1,1}-d_{n',3,1}) + (t_{n,1,1}- t_{n',3,1})\, \nonumber \\
\end{align}

where $\Delta^0$ stands for $\sigma_{0,n,1}^{0}-\sigma_{0,n',3}^{0}$, the difference between
the eigenfrequencies without rotation.

For high frequency p-modes, the radial contributions are concentrated
in the outer layers and  are then nearly the  same  for  $(n,\ell=1)$ and  $(n',\ell=3)$. 
It is then legitimate to neglect
the differences in their radial contributions. We then omit the radial order 
subscripts $n,n'$. For the same reason, the Ledoux constants are also quite similar $\beta_1 \sim \beta_3$.  Then
\begin{align}
\Delta_1 \, \approx  \Delta^0 \, &  +  (d_{1,1}-d_{3,1})+ 
 \,  (t_{1,1}- t_{3,1}) \, \nonumber
\end{align}
Similarly 
\begin{align}
\Delta_{-1} \,\approx  \, \Delta^0 \, & + \,   (d_{n,1,1}-d_{3,1})\, -  \,  (t_{1,1}- t_{3,1}) \nonumber
\end{align}
Hence, the quantity $(\Delta_{1}\, - \, \Delta_{-1} )$ in Eq.(\ref{A_splitND_init}):
\begin{align}
\Delta_{-1} \,- \,  \Delta_{1} \approx &  2 \, \,(t_{1,1}- t_{3,1}) 
\end{align}
is of third order. One can then approximate $\Delta_{-1} \,\sim \,  \Delta_{1}$  
in the square root of Eq.(\ref{A_splitND_init}), and the correction due to near degeneracy  is given by:
 \begin{align}
\rm S_{\ell=1}^{deg} \, - \, \rm S_{\ell=1}^{ND} \, 
\approx & \, \,  (t_{1,1}- t_{3,1}) \nonumber \\ 
&+ \, \frac{ \sqrt{\Delta_1^2 \,+ \, \mathcal{H}_1^2} \, - \,
 \sqrt{\Delta_{1}^2 \,+ \, \mathcal{H}_{-1}^2} }{2} 
\end{align}
For the coupling term, one can separate the second $\mathcal{H}_2$  and 
third $\mathcal{H}_3$ order terms as 
$\mathcal{H}_m\, = \, \mathcal{H}_2  \, + \, m \, \mathcal{H}_3 $.
Thus $\mathcal{H}_1\, = \, \mathcal{H}_2  \, + \, \mathcal{H}_3 $, 
 and $\mathcal{H}_{-1}\, = \, \mathcal{H}_2  \, - \, \mathcal{H}_3 $.
With $|\mathcal{H}_3| <<|\mathcal{H}_2|$,   the near degeneracy 
correction to the splitting becomes: 

\begin{align}
\label{A_splitdeg_fin}
\rm S_{\ell=1}^{deg} \,- \, \rm S_{\ell=1}^{ND} \, \approx \,   (t_{1,1}- t_{3,1})\, + \, \frac{2 \mathcal{H}_2 \mathcal{H}_3}{\sqrt{\Delta_{1}^2 \, + \, \mathcal{H}_2^2} }
\end{align}
which shows that for slow rotators, such as HD 181420,
if cubic order effects are neglected, then the near degeneracy 
contribution is zero, and the splitting is a linear function of rotation. 
If cubic order effects are included, then near degeneracy corrections affect the splitting, and the  departure from a linear splitting is dominated by distorsion (predominantly in $\mathcal{H}_2$) .  

\section{Contribution of the latitdinal shear to the splittings}
\label{app_shear}
In order to be able to compute the splittings from Eq.(\ref{Sm}) and Eq.(\ref{sp}), one must specify a rotation
law. It is convenient to assume the following form:
\begin{equation}
\Omega(\rm r,\theta)=  \sum_{\rm s=0}^{\rm s_{max}} ~ \Omega_{\rm 2s}(\rm r) ~(\cos\theta)^{\rm 2s}
\end{equation}
where $\theta$ is the colatitude and we have limited our investigation $s_{max}=1$. The surface rotation rate at the equator is $\Omega(\rm r= \rm R,\theta=\pi/2) =\Omega_0(r\rm =\rm R)$.

Inserting Eq.(\ref{latit}) into Eq.(\ref{sp}) 
yields the following  expression for the generalized splitting:
 \begin{eqnarray}
\rm S_{\rm m}  &=&  \frac{1}{\Omega_k} \, \int_0^{\rm R}   \Omega_0(\rm r) ~\rm K(\rm r) ~\rm dr \, + \, \frac{1}{\Omega_k} \,\sum_{\rm s=0}^{\rm s=2}  \rm m^{\rm 2s} ~ \rm H_{\rm m,s}(\Omega) 
 \end{eqnarray}
 with \citep[see][]{Goupil2010},
\begin{equation}
\rm H_{\rm m,s} (\Omega) = -
{1 / \rm I}  \int_0^{\rm R}   \Bigl[ \rm R_{\rm s} ~\Bigl(\xi^2_{\rm r}  -2\xi_{\rm r} \xi_{\rm h}+ \xi^2_{\rm h}   (\Lambda-1)\Bigr) +\rm Q_{\rm s} ~ \xi^2_{\rm h}\Bigr)   \Bigr]  \rho_0 \rm r^2
dr \nonumber 
 \end{equation} 
 \ni
where $\rm R_{\rm s}$ and $\rm Q_{\rm s}$ depend on $\Omega_2,\Omega_4$ and $\Lambda=\ell(\ell+1)$. \\
For example, if shellular rotation is expected, then $\Omega(\rm r,\theta)= \Omega_0(\rm r)$ and $\rm s_{\rm max}=0$. Moreover, $\Omega_2, \, \Omega_4=0$, i.e.  $\rm R_{\rm j}= \rm Q_{\rm j}=0$  for $\rm j=0,2$, and:
 \begin{eqnarray}
\rm S_{\rm m}  &=& - {1  / \rm I}      \int_0^{\rm R} ~ \frac{\Omega_{0}(\rm r)}{\Omega_k}~  \Bigl[  \Bigr.
\xi_{\rm r}^2   - 2\xi_{\rm r} \xi_{\rm h}     + \Lambda ~ \xi_{\rm h}^2 \Bigl.  \Bigr] ~\rho_0 \rm r^2 \rm dr  
\end{eqnarray}
\medskip
If we consider latitudinally differential rotation only, $\Omega_{2j}, \rm j=0,2$ are depth independent, $\Omega_0=\Omega_{\rm equator}$, $\Omega_2 = -\Delta \Omega$, and $\Omega_4=0$.  $\rm R_{\rm s}$ and $\rm Q_{\rm s}$ are constant and:
\begin{eqnarray}
\rm S_{\rm m}  =   \frac{\Omega_{0}}{\Omega_k} ~ \beta  +  \frac{1}{\Omega_k} ~\sum_{\rm s=0}^{\rm s=1}  \rm m^{\rm 2s} ~  (\rm R_{\rm s}(\Omega) ~\beta + \rm Q_{\rm s}(\Omega)~ \gamma )
\end{eqnarray}
with $\beta$ and $\gamma$ defined as:
\begin{eqnarray}
\label{def_beta}
&& \beta = \frac{1}{\rm I} \int_0^{\rm R} \left[  \xi_{\rm r}^2-2 \xi_{\rm r} \xi_{\rm h} + (\Lambda -1) \xi_{\rm h}^2 \right]  \rho_0 \rm r^2 \rm dr \\
&& \gamma= - \frac{1}{\rm I} \int_0^{\rm R} ~ \xi^2_{\rm h} ~ \rho_0 \rm r^2 ~\rm dr  
\end{eqnarray}
\ni
and I being the inertia of the mode:
\begin{equation}
\rm I = \int_0^{\rm R} ~ (\xi_{\rm r}^2+\Lambda \xi^2_{\rm h}) ~ \rho_0 \rm r^2 ~\rm dr 
\end{equation}
Then the rotational splitting of a $\{\rm n,\ell,\rm m\}$ mode is given by

\begin{equation}
\rm S_{\ell,\rm m}=\frac{1}{\Omega_k}\, \left[ \Omega_{0} \, \beta + \rm R_0^{\ell} \, \beta + \rm Q_0^{\ell} \, \gamma + \rm m^2 \left( \rm R_1^{\ell} \, \beta + \rm Q_1^{\ell} \, \gamma \right)\right]
\end{equation}
In \cite{Goupil2010}, $\rm Q_{\rm s}$ and $\rm R_{\rm s}$ are given for $\rm s = 0,1,2$:
\begin{align}
\rm R_0^{\ell} =& - \frac{\Delta \Omega}{\Omega_0} ~ \frac{2 \Lambda-1}{4 \Lambda -3}  \\
\rm R_1^{\ell} =& \, \frac{\Delta \Omega}{\Omega_0}  ~ \frac{2}{4 \Lambda -3} \\
\rm Q_0^{\ell} =&  -\frac{\Delta \Omega }{\Omega_0} ~ \frac{2 (1-3\Lambda)}{4 \Lambda -3} \\
\rm Q_1^{\ell} =&   -\frac{\Delta \Omega}{\Omega_0}  ~ \frac{10}{4 \Lambda -3}
\end{align}
\ni
For $\ell=1$ and $\ell=2$ modes, this yields:
\begin{itemize}
\item[a.] for $\ell=1$, $\rm m=\pm 1$  (i.e. $\Lambda=2$): 
 \begin{align}
\rm S_{1,1}  = \,  \frac{\Omega_{0}}{\Omega_k} ~ \beta ~ \left[ 1 - \frac{1}{5}~ \frac{\Delta \Omega}{\Omega_{0}} \right]
\label{S11_app}
 \end{align}
\item[b.] for $\ell=2$, $m=\pm1$ (i.e. $\Lambda=6$) : 
 \begin{align}
\rm S_{2,1}  =  \, \frac{\Omega_{0}}{\Omega_k} ~ \left[\beta ~ + \,\frac{1}{7}\,  \frac{\Delta \Omega}{\Omega_{0}} ~ ( 8 ~ \gamma - 3 ~ \beta) \right]
 \end{align}
\item[c.] for $\ell=2$, $m=\pm2$ (i.e. $\Lambda=6$) : 
 \begin{align}
\rm S_{2,2}  = \, \frac{\Omega_{0}}{\Omega_k} \, \left[  2~ \beta ~ - \frac{1}{7}\,  \frac{\Delta \Omega}{\Omega_{0}} ~ ( 2 ~ \gamma +~ \beta)\right]
\label{S21_app}
 \end{align}
\end{itemize}

\section{Conditions for significant near degeneracy}
\label{App_HmDm}

As already mentionned, near degeneracy  between two modes  occurs whenever 
their frequencies are close, their azimuthal numbers are equal, and their angular degrees differ by 2. However the magnitude of the near-degenerate corrections to both frequencies also depends on the magnitude of the coupling term $H_m$. In turn 
the magnitude of $H_m$ depends on the nature of the involved modes whether they are g-modes,
p-modes or mixed modes with a dominant p or g nature.
  
The g-mode spectrum is much denser than the p-mode one.  Hence, 
as shown in Fig. C.2, $\Delta_m$ is much smaller than for p-modes. This ought to favor 
near degeneracy. However, the coupling term $H_m$  for g-modes 
is much smaller than for p-modes. This is due to
the fact that distorsion effects are small for g-modes and therefore the overall
(second  and third order)  contribution to $H_m$ remains small. As a result, the coupling
term is much smaller than $\Delta_m$ for g-modes, which are then hardly coupled.
\begin{figure}[h!]
\begin{center}
\includegraphics[scale=0.33,angle=-90]{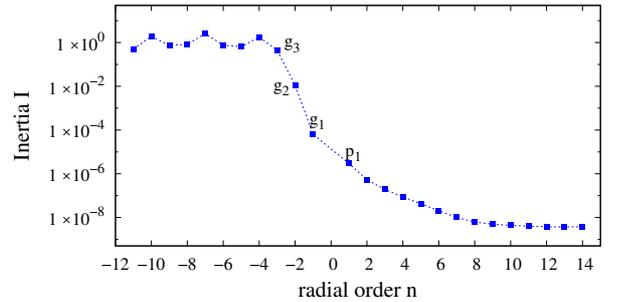}
\caption{\label{inertiaplot}
Inertia of axisymmetric $\ell=1$ modes as a function of to the radial order, n, for an evolved 8.5 M$_{\odot}$ $\beta$ Cephei model, uniformly rotating at $30\%\, \Omega_k$,i.e. $160 km.s^{-1}$ (see Table \ref{tabBceph}, Sect.4.1).}
\end{center}
\end{figure}
\begin{figure}[h!]
\begin{center}
\includegraphics[scale=0.33,angle=-90]{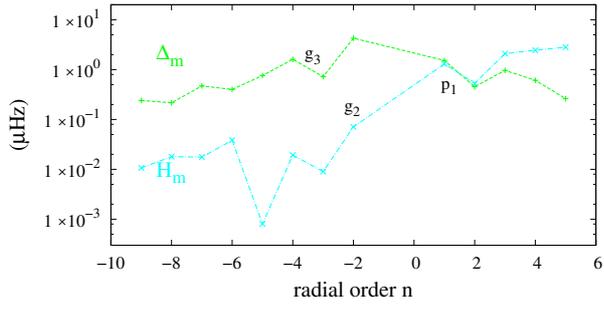}
\caption{\label{Hmplot} The terms $\Delta_m$ and $\mathcal{H}_m$ which 
are involved in near degeneracy corrections (see Eq.(\ref{ND_correction}))  for $\ell=1$ modes
computed for the same $\beta$ Cephei model as in Fig.C.1. A logscale is used for the y axis.}
\end{center}
\end{figure}

\onecolumn

\section{Values for third order contributions to the frequencies and to the splittings}
\begin{center}
\begin{table*}[h!]
\caption{Scaled contributions to the splittings of $\ell=1$ g-modes due to: implicit
third order terms in the eigenfrequency ($T^{eigen}$), Coriolis effects ($T^T$),
distorsion ($T^D$), and  coupling of the two ($T^C$). The impact of near degeneracy is so small that its 
contribution is fully negligible and is therefore not listed. 
The first row lists the radial order and the
second row the centroid mode $m=0$  frequency.}             % title of Table
\label{TabT3_g} 
\centering                          % used for centering table
\vspace*{1.5cm}\begin{tabular}{c c c c c c}        % centered columns (4 columns)
\hline\hline                 % inserts double horizontal lines
n & $\sigma_0$  & $T_3^{eigen} $ &$T_3^T $ &$T_3^D $ &$T_3^C $  \\    % table heading 
\hline    
$  -8$ &  \,\, $   0.25$ &  \,\, $ 0.672E+02$ &  \,\,$ 0.529E+04$ &  \,\,$-0.390E+01$ &  \,\,$-0.557E+03$ \\
$  -7$ &  \,\, $   0.29$ &  \,\, $ 0.705E+02$ &  \,\,$ 0.391E+04$ &  \,\,$ 0.214E+02$ &  \,\,$-0.424E+03$ \\
$  -6$ &  \,\, $   0.34$ &  \,\, $ 0.770E+02$ &  \,\,$ 0.285E+04$ &  \,\,$ 0.379E+02$ &  \,\,$-0.362E+03$ \\
$  -5$ &  \,\, $   0.41$ &  \,\, $ 0.943E+02$ &  \,\,$ 0.208E+04$ &  \,\,$ 0.518E+02$ &  \,\,$-0.354E+03$ \\
$  -4$ &  \,\, $   0.52$ &  \,\, $ 0.123E+03$ &  \,\,$ 0.153E+04$ &  \,\,$ 0.683E+02$ &  \,\,$-0.371E+03$ \\
$  -3$ &  \,\, $   0.70$ &  \,\, $ 0.155E+03$ &  \,\,$ 0.111E+04$ &  \,\,$ 0.905E+02$ &  \,\,$-0.374E+03$ \\
$  -2$ &  \,\, $   1.06$ &  \,\, $ 0.850E+02$ &  \,\,$ 0.776E+03$ &  \,\,$ 0.116E+03$ &  \,\,$-0.318E+03$ \\
$  -1$ &  \,\, $   3.52$ &  \,\, $-0.269E+04$ &  \,\,$ 0.833E+03$ &  \,\,$-0.439E+03$ &  \,\,$ 0.851E+03$ \\
\hline                                   %inserts single line

\end{tabular}
\end{table*}

\begin{table*}[h!]
\caption{Scaled contributions to the splittings of $\ell=1$ p-modes due to third order effects divided by
 the square of the central mode frequency. The first row lists the radial order and the
  second row the centroid mode $m=0$ frequency.}             % title of Table
\label{TabT3_p} 
\centering                          % used for centering table
\vspace*{1.5cm}\begin{tabular}{c c c c c c c}        % centered columns (4 columns)
\hline\hline                 % inserts double horizontal lines
n & $\sigma_0$ & $T_3^{eigen} / \sigma_0^2$ &$T_3^T/ \sigma_0^2$ &$T_3^D/ \sigma_0^2$ &$T_3^C/ \sigma_0^2$ &$T_3^{deg}/ \sigma_0^2$ \\    % table heading 
\hline     
$  -1$ &  \,\, $   3.52$ &  \,\, $-0.217E+03$ &  \,\,$ 0.672E+02$ &  \,\,$-0.355E+02$ &  \,\,$ 0.687E+02$ &  \,\,$-0.782E-05$ \\
$   1$ &  \,\, $   4.73$ &  \,\, $-0.140E+03$ &  \,\,$ 0.409E+02$ &  \,\,$-0.181E+02$ &  \,\,$ 0.433E+02$ &  \,\,$ 0.697E+00$ \\
$   2$ &  \,\, $   5.78$ &  \,\, $-0.106E+03$ &  \,\,$ 0.285E+02$ &  \,\,$-0.115E+02$ &  \,\,$ 0.343E+02$ &  \,\,$ 0.174E+02$ \\
$   3$ &  \,\, $   6.85$ &  \,\, $-0.862E+02$ &  \,\,$ 0.181E+02$ &  \,\,$-0.554E+01$ &  \,\,$ 0.264E+02$ &  \,\,$ 0.318E+02$ \\
$   4$ &  \,\, $   7.96$ &  \,\, $-0.727E+02$ &  \,\,$ 0.112E+02$ &  \,\,$-0.666E+00$ &  \,\,$ 0.192E+02$ &  \,\,$ 0.380E+02$ \\
$   5$ &  \,\, $   9.11$ &  \,\, $-0.636E+02$ &  \,\,$ 0.767E+01$ &  \,\,$ 0.217E+01$ &  \,\,$ 0.145E+02$ &  \,\,$ 0.435E+02$ \\
$   6$ &  \,\, $  10.30$ &  \,\, $-0.566E+02$ &  \,\,$ 0.581E+01$ &  \,\,$ 0.384E+01$ &  \,\,$ 0.110E+02$ &  \,\,$ 0.447E+02$ \\
$   7$ &  \,\, $  11.49$ &  \,\, $-0.514E+02$ &  \,\,$ 0.494E+01$ &  \,\,$ 0.456E+01$ &  \,\,$ 0.884E+01$ &  \,\,$ 0.472E+02$ \\
$   8$ &  \,\, $  12.67$ &  \,\, $-0.477E+02$ &  \,\,$ 0.440E+01$ &  \,\,$ 0.472E+01$ &  \,\,$ 0.762E+01$ &  \,\,$ 0.503E+02$ \\
$   9$ &  \,\, $  13.82$ &  \,\, $-0.447E+02$ &  \,\,$ 0.374E+01$ &  \,\,$ 0.471E+01$ &  \,\,$ 0.688E+01$ &  \,\,$ 0.540E+02$ \\
$  10$ &  \,\, $  14.96$ &  \,\, $-0.417E+02$ &  \,\,$ 0.290E+01$ &  \,\,$ 0.489E+01$ &  \,\,$ 0.601E+01$ &  \,\,$ 0.555E+02$ \\
$  11$ &  \,\, $  16.11$ &  \,\, $-0.387E+02$ &  \,\,$ 0.221E+01$ &  \,\,$ 0.523E+01$ &  \,\,$ 0.494E+01$ &  \,\,$ 0.543E+02$ \\
$  12$ &  \,\, $  17.28$ &  \,\, $-0.360E+02$ &  \,\,$ 0.181E+01$ &  \,\,$ 0.552E+01$ &  \,\,$ 0.395E+01$ &  \,\,$ 0.527E+02$ \\
$  13$ &  \,\, $  18.46$ &  \,\, $-0.338E+02$ &  \,\,$ 0.161E+01$ &  \,\,$ 0.570E+01$ &  \,\,$ 0.316E+01$ &  \,\,$ 0.516E+02$ \\
$  14$ &  \,\, $  19.63$ &  \,\, $-0.318E+02$ &  \,\,$ 0.149E+01$ &  \,\,$ 0.580E+01$ &  \,\,$ 0.252E+01$ &  \,\,$ 0.508E+02$ \\
$  15$ &  \,\, $  20.81$ &  \,\, $-0.301E+02$ &  \,\,$ 0.140E+01$ &  \,\,$ 0.585E+01$ &  \,\,$ 0.196E+01$ &  \,\,$ 0.498E+02$ \\
\hline                                   %inserts single line
\end{tabular}
\end{table*}
\end{center}

\vspace*{1.5cm}
\begin{table*}[h!]
\centering                          % used for centering table
\caption{Different order contributions to the mode frequencies for various radial orders, n. All contributions are scaled by $\Omega_k$.}             % title of Table
\label{TabSig} 
\vspace*{1.5cm}
\begin{tabular}{c c c c c c c c c c c}        % centered columns (4 columns)
\hline\hline                 % inserts double horizontal lines
n & $\sigma_0$ & $\sigma_1$ & $\sigma_2^{eigen}$ &$\sigma_2^T$ &$\sigma_2^D$  \\    % table heading 
\hline                        % inserts single horizontal line
$  -8$ &  \,\, $   0.25$ &  \,\, $ 0.777E-01$ &  \,\, $ 0.989E-02$ &  \,\, $ 0.268E-01$ & \,\,  $-0.167E-03$\\
$  -7$ &  \,\, $   0.29$ &  \,\, $ 0.782E-01$ &  \,\, $ 0.857E-02$ &  \,\, $ 0.238E-01$ & \,\,  $-0.189E-03$\\
$  -6$ &  \,\, $   0.34$ &  \,\, $ 0.790E-01$ &  \,\, $ 0.713E-02$ &  \,\, $ 0.206E-01$ & \,\,  $-0.228E-03$\\
$  -5$ &  \,\, $   0.41$ &  \,\, $ 0.802E-01$ &  \,\, $ 0.562E-02$ &  \,\, $ 0.173E-01$ & \,\,  $-0.293E-03$\\
$  -4$ &  \,\, $   0.52$ &  \,\, $ 0.819E-01$ &  \,\, $ 0.407E-02$ &  \,\, $ 0.141E-01$ & \,\,  $-0.392E-03$\\
$  -3$ &  \,\, $   0.70$ &  \,\, $ 0.839E-01$ &  \,\, $ 0.252E-02$ &  \,\, $ 0.108E-01$ & \,\,  $-0.529E-03$\\
$  -2$ &  \,\, $   1.06$ &  \,\, $ 0.859E-01$ &  \,\, $ 0.954E-03$ &  \,\, $ 0.737E-02$ & \,\,  $-0.729E-03$\\
$  -1$ &  \,\, $   3.52$ &  \,\, $ 0.142E+00$ &  \,\, $ 0.677E-02$ &  \,\, $ 0.325E-02$ & \,\,  $-0.101E-01$\\
$   1$ &  \,\, $   4.73$ &  \,\, $ 0.141E+00$ &  \,\, $ 0.497E-02$ &  \,\, $ 0.247E-02$ & \,\,  $-0.143E-01$\\
$   2$ &  \,\, $   5.78$ &  \,\, $ 0.141E+00$ &  \,\, $ 0.401E-02$ &  \,\, $ 0.205E-02$ & \,\,  $-0.188E-01$\\
$   3$ &  \,\, $   6.85$ &  \,\, $ 0.142E+00$ &  \,\, $ 0.337E-02$ &  \,\, $ 0.176E-02$ & \,\,  $-0.224E-01$\\
$   4$ &  \,\, $   7.96$ &  \,\, $ 0.142E+00$ &  \,\, $ 0.288E-02$ &  \,\, $ 0.153E-02$ & \,\,  $-0.253E-01$\\
$   5$ &  \,\, $   9.11$ &  \,\, $ 0.143E+00$ &  \,\, $ 0.251E-02$ &  \,\, $ 0.135E-02$ & \,\,  $-0.286E-01$\\
$   6$ &  \,\, $  10.30$ &  \,\, $ 0.144E+00$ &  \,\, $ 0.221E-02$ &  \,\, $ 0.121E-02$ & \,\,  $-0.319E-01$\\
$   7$ &  \,\, $  11.49$ &  \,\, $ 0.145E+00$ &  \,\, $ 0.197E-02$ &  \,\, $ 0.109E-02$ & \,\,  $-0.356E-01$\\
$   8$ &  \,\, $  12.67$ &  \,\, $ 0.145E+00$ &  \,\, $ 0.178E-02$ &  \,\, $ 0.997E-03$ & \,\,  $-0.397E-01$\\
$   9$ &  \,\, $  13.82$ &  \,\, $ 0.146E+00$ &  \,\, $ 0.163E-02$ &  \,\, $ 0.919E-03$ & \,\,  $-0.438E-01$\\
$  10$ &  \,\, $  14.96$ &  \,\, $ 0.146E+00$ &  \,\, $ 0.150E-02$ &  \,\, $ 0.852E-03$ & \,\,  $-0.471E-01$\\
$  11$ &  \,\, $  16.11$ &  \,\, $ 0.146E+00$ &  \,\, $ 0.139E-02$ &  \,\, $ 0.794E-03$ & \,\,  $-0.499E-01$\\
$  12$ &  \,\, $  17.28$ &  \,\, $ 0.147E+00$ &  \,\, $ 0.129E-02$ &  \,\, $ 0.743E-03$ & \,\,  $-0.526E-01$\\
$  13$ &  \,\, $  18.46$ &  \,\, $ 0.147E+00$ &  \,\, $ 0.121E-02$ &  \,\, $ 0.697E-03$ & \,\,  $-0.557E-01$\\
$  14$ &  \,\, $  19.63$ &  \,\, $ 0.147E+00$ &  \,\, $ 0.114E-02$ &  \,\, $ 0.657E-03$ & \,\,  $-0.588E-01$\\
$  15$ &  \,\, $  20.81$ &  \,\, $ 0.147E+00$ &  \,\, $ 0.107E-02$ &  \,\, $ 0.621E-03$ & \,\,  $-0.619E-01$\\
\hline                        % inserts single horizontal line
\end{tabular}
\newline
\vspace*{1.5cm}
\newline
\begin{tabular}{c c c c c c c c c c c}        % centered columns (4 columns)
\hline\hline                 % inserts double horizontal lines
n & $\sigma_3^{eigen}$ &$\sigma_3^T$ &$\sigma_3^D$ &$\sigma_3^C$ &$\sigma^{\rm deg} - \sigma^{\rm no deg}$\\    % table heading 
\hline                        % inserts single horizontal line
$  -8$ &  \,\, $ 0.186E-04$ &  \,\,$ 0.193E-02$ &  \,\,$ 0.199E-04$ &  \,\,$-0.528E-03$ &  \,\,$< 10^{-5}$ \\
$  -7$ &  \,\, $ 0.259E-04$ &  \,\,$ 0.172E-02$ &  \,\,$ 0.325E-04$ &  \,\,$-0.439E-03$ &  \,\,$< 10^{-5}$ \\
$  -6$ &  \,\, $ 0.185E-04$ &  \,\,$ 0.151E-02$ &  \,\,$ 0.423E-04$ &  \,\,$-0.394E-03$ &  \,\,$< 10^{-5}$ \\
$  -5$ &  \,\, $ 0.582E-05$ &  \,\,$ 0.130E-02$ &  \,\,$ 0.515E-04$ &  \,\,$-0.378E-03$ &  \,\,$< 10^{-5}$ \\
$  -4$ &  \,\, $-0.205E-04$ &  \,\,$ 0.105E-02$ &  \,\,$ 0.607E-04$ &  \,\,$-0.362E-03$ &  \,\,$< 10^{-5}$ \\
$  -3$ &  \,\, $-0.696E-04$ &  \,\,$ 0.767E-03$ &  \,\,$ 0.677E-04$ &  \,\,$-0.313E-03$ &  \,\,$< 10^{-5}$ \\
$  -2$ &  \,\, $-0.178E-03$ &  \,\,$ 0.429E-03$ &  \,\,$ 0.640E-04$ &  \,\,$-0.207E-03$ &  \,\,$< 10^{-5}$ \\
$  -1$ &  \,\, $ 0.139E-03$ &  \,\,$ 0.173E-03$ &  \,\,$-0.960E-04$ &  \,\,$ 0.187E-03$ &  \,\,$ 0.507E-03$ \\
$   1$ &  \,\, $ 0.883E-04$ &  \,\,$ 0.144E-03$ &  \,\,$-0.655E-04$ &  \,\,$ 0.156E-03$ &  \,\,$ 0.100E-02$ \\
$   2$ &  \,\, $ 0.635E-04$ &  \,\,$ 0.123E-03$ &  \,\,$-0.506E-04$ &  \,\,$ 0.151E-03$ &  \,\,$ 0.262E-02$ \\
$   3$ &  \,\, $ 0.352E-04$ &  \,\,$ 0.930E-04$ &  \,\,$-0.288E-04$ &  \,\,$ 0.137E-03$ &  \,\,$ 0.447E-02$ \\
$   4$ &  \,\, $ 0.496E-05$ &  \,\,$ 0.670E-04$ &  \,\,$-0.402E-05$ &  \,\,$ 0.116E-03$ &  \,\,$ 0.596E-02$ \\
$   5$ &  \,\, $-0.192E-04$ &  \,\,$ 0.526E-04$ &  \,\,$ 0.149E-04$ &  \,\,$ 0.998E-04$ &  \,\,$ 0.203E-02$ \\
$   6$ &  \,\, $-0.359E-04$ &  \,\,$ 0.450E-04$ &  \,\,$ 0.299E-04$ &  \,\,$ 0.854E-04$ &  \,\,$ 0.299E-02$ \\
$   7$ &  \,\, $-0.489E-04$ &  \,\,$ 0.428E-04$ &  \,\,$ 0.395E-04$ &  \,\,$ 0.767E-04$ &  \,\,$ 0.415E-02$ \\
$   8$ &  \,\, $-0.595E-04$ &  \,\,$ 0.420E-04$ &  \,\,$ 0.452E-04$ &  \,\,$ 0.729E-04$ &  \,\,$ 0.187E-01$ \\
$   9$ &  \,\, $-0.713E-04$ &  \,\,$ 0.390E-04$ &  \,\,$ 0.492E-04$ &  \,\,$ 0.718E-04$ &  \,\,$ 0.244E-01$ \\
$  10$ &  \,\, $-0.830E-04$ &  \,\,$ 0.327E-04$ &  \,\,$ 0.553E-04$ &  \,\,$ 0.679E-04$ &  \,\,$ 0.295E-01$ \\
$  11$ &  \,\, $-0.906E-04$ &  \,\,$ 0.269E-04$ &  \,\,$ 0.637E-04$ &  \,\,$ 0.601E-04$ &  \,\,$ 0.336E-01$ \\
$  12$ &  \,\, $-0.964E-04$ &  \,\,$ 0.237E-04$ &  \,\,$ 0.721E-04$ &  \,\,$ 0.516E-04$ &  \,\,$ 0.379E-01$ \\
$  13$ &  \,\, $-0.101E-03$ &  \,\,$ 0.225E-04$ &  \,\,$ 0.796E-04$ &  \,\,$ 0.441E-04$ &  \,\,$ 0.434E-01$ \\
$  14$ &  \,\, $-0.105E-03$ &  \,\,$ 0.222E-04$ &  \,\,$ 0.862E-04$ &  \,\,$ 0.374E-04$ &  \,\,$ 0.497E-01$ \\
$  15$ &  \,\, $-0.110E-03$ &  \,\,$ 0.221E-04$ &  \,\,$ 0.922E-04$ &  \,\,$ 0.309E-04$ &  \,\,$ 0.564E-01$ \\
\hline                        % inserts single horizontal line
\end{tabular}\end{table*}

\end{appendix}

\twocolumn

\end{document}